\documentclass[12pt, preprint]{emulateapj}
\usepackage[]{natbib}
\shorttitle{GC color magnitude relations}
\shortauthors{Mieske et al.}
\def\etal{{\it et~al.}~}

\def\kmm{{\tt KMM~}}

\begin{document}

\title{The ACS Fornax Cluster Survey. IX. The Color-Magnitude
Relation of Globular Cluster Systems\altaffilmark{1}}

\author{Steffen Mieske\altaffilmark{2}, Andr\'es
Jord\'an\altaffilmark{3,4}, Patrick C\^{o}t\'{e}\altaffilmark{5}, Eric
W. Peng\altaffilmark{6}, Laura Ferrarese\altaffilmark{5}, John
P. Blakeslee\altaffilmark{5}, Simona Mei\altaffilmark{7,8}, Holger Baumgardt\altaffilmark{9}, John
L. Tonry\altaffilmark{10}, Leopoldo Infante\altaffilmark{3}, \& Michael J. West\altaffilmark{2}}
\altaffiltext{1}{Based on observations with the NASA/ESA {\it Hubble
Space Telescope} obtained at the Space Telescope Science Institute,
which is operated by the association of Universities for Research in
Astronomy, Inc., under NASA contract NAS 5-26555.}
\altaffiltext{2}{European Southern Observatory,
Alonso de Cordova 3107, Vitacura, Santiago, Chile}
\altaffiltext{3}{Departamento de Astronomía y Astrofísica, Pontificia Universidad Católica de Chile, Santiago 22, Chile}
\altaffiltext{4}{Harvard-Smithsonian Center for Astrophysics, Cambridge, MA 02138, USA}
\altaffiltext{5}{Herzberg Institute of Astrophysics, Victoria,  BC V9E 2E7, Canada} 
\altaffiltext{6}{Department of Astronomy, Peking University, Beijing 100871, China}
\altaffiltext{7}{University of Paris Denis Diderot, 75205 Paris Cedex 13, France}
\altaffiltext{8}{GEPI, Observatoire de Paris, Section de Meudon, 5 Place J. Janssen, 92195 Meudon Cedex, France}
\altaffiltext{9}{Argelander Institut f\"ur Astronomie, Auf dem H\"ugel 71, 53121 Bonn, Germany}
\altaffiltext{10}{Institute for Astronomy,
University of Hawaii, 2680 Woodlawn Drive, Honolulu, HI 96822} 

\begin{abstract}
\noindent We investigate the color-magnitude relation for globular
clusters (GCs) --- the so-called ``blue tilt" --- detected in the {\it
ACS Fornax Cluster Survey} and using the combined sample of GCs from
the {\it ACS Fornax and Virgo Cluster Surveys}. We find a tilt of
$\gamma_z \equiv {d(g-z)}/{dz}=-0.0257 \pm 0.0050$ for the full GC
sample of the Fornax Cluster Survey ($\approx$5800 GCs). This is
slightly shallower than the value $\gamma_z=-0.0459 \pm 0.0048$ found
for the Virgo Cluster Survey GC sample ($\approx$11100 GCs).  The
slope for the merged Fornax and Virgo datasets ($\approx$16900 GCs) is
$\gamma_z=-0.0293 \pm 0.0085$, corresponding to a mass-metallicity
relation of $Z \propto M^{\rm 0.43 \pm 0.12}$. We find that the blue
tilt sets in at masses in excess of $M \sim 2 \times 10^5
M_{\odot}$. The tilt is stronger for GCs belonging to high-mass
galaxies ($M_* \gtrsim 5\times10^{10}M_{\odot}$) than for those in
low-mass galaxies ($M_* \lesssim 5\times10^{10}M_{\odot}$). It is also
more pronounced for GCs with smaller galactocentric distances. Our
findings suggest a range of mass-metallicity relations $Z_{\rm GC}
\propto M_{\rm GC}^{0.3-0.7}$ which vary as a function of host galaxy
mass/luminosity, a scaling similar to that observed for dwarf
spheroidal galaxies.  We compare our observations to a recent model of
star cluster self-enrichment with generally favorable results. We suggest
that, within the context of this model, the proto-cluster clouds out
of which the GCs formed may have had density profiles slightly steeper
than isothermal and/or star formation efficiencies somewhat below
0.3. We caution, however, that the significantly different appearance
of the CMDs defined by the GC systems associated with galaxies of
similar mass and morphological type pose a challenge to any single
mechanism, including self-enrichment, that seeks to explain
generically the observed GC color-magnitude relations. We therefore
suggest that the detailed (and stochastic) merger/accretion histories
of individual galaxies have likely played a non-negligible role
determining the distribution of GCs in the color-magnitude diagrams of
individual GC systems.

\end{abstract}


\section{Introduction}
In recent years, deep space-based imaging, in combination with
high-resolution spectroscopy of individual stars, has revealed the
existence of multiple stellar populations in several  globular
clusters (GCs) belonging to the Milky Way (e.g., Lee {\it et al.}~\cite{Lee99}, Bedin et
al.`\cite{Bedin04}, Piotto {\it et al.}~\cite{Piotto05} \& ~\cite{Piotto07},
Villanova {\it et al.}~\cite{Villan07}, Milone {\it et al.}~\cite{Milone08}). This
finding has both undermined the long-held view that star clusters are
simple systems consisting of single-age, single-metallicity stellar
populations, and renewed interest in GC self-enrichment
scenarios (e.g., Frank \& Gisler~\cite{Frank76}, Smith~\cite{Smith96},
Gnedin \etal~\cite{Gnedin02}, Parmentier \& Gilmore~\cite{Parmen01},
Dopita \& Smith~\cite{Dopita86}, Morgan \& Lake~\cite{Morgan89}, Thoul
\etal~\cite{Thoul02}, Parmentier~\cite{Parmen04}, Recchi \&
Danziger~\cite{Recchi05}, Caloi \& D'Antona~\cite{Caloi07}, Bailin \&
Harris~\cite{Bailin09}).

As a complement to such studies, observations of (unresolved) GCs in
galaxies beyond the Local Group offer the advantage of much larger and
more homogeneous databases of GC colors and magnitudes with which to
investigate overall trends between integrated GC properties and the
properties of the GC systems as a whole. Observations from HST, using
both WFPC2 and ACS, have been pivotal in this area (e.g., C\^ot\'e et
al.~\cite{Cote04}, Harris {\it et al.}~\cite{Harris06} \& \cite{Harris09},
Peng {\it et al.}~\cite{Peng06}, Jordan {\it et al.}~\cite{Jordan07a}).  One of
the more surprising results to have emerged from these studies has
been the discovery of a relationship between the magnitudes and colors
of {\it individual} GCs, in the sense that GCs associated with the
blue subpopulation become progressively redder at high luminosities
(Harris {\it et al.}~\cite{Harris06}, Strader {\it et al.}~\cite{Strade06}, Mieske
{\it et al.}~\cite{Mieske06}, Spitler {\it et al.}~\cite{Spitle06},
Humphrey~\cite{Humphr09}, Cockcroft {\it et al.}~\cite{Cockcr09}, Harris et
al.~\cite{Harris09b}, Forbes et al.~\cite{Forbes10}).

This relation, which is now commonly refered to as the ``blue tilt",
has also been detected in ground-based imaging of extragalactic GC
systems (Forte {\it et al.}~\cite{Forte07}, Wehner {\it et al.}~\cite{Wehner08}. Harris~\cite{Harris09c})
and possibly for M31 as well (Fan, Ma \& Zhou~\cite{Fan09}). If
interpreted as a mass-metallicity relation, the tilt corresponds to a
scaling with mass of $Z \propto M^{0.3-0.7}$. This is comparable to
the relation $Z \propto M^{0.6-0.7}$ defined by the faintest dwarf
spheroidal galaxies of the Milky Way (e.g. Simon \&
Geha~\cite{Simon07}, Kirby {\it et al.}~\cite{Kirby08}) which have luminous
masses comparable to GCs but gravitating masses 1-2 order of magnitude
larger. Recently, Mieske {\it et al.}~\cite{Mieske06} and Blakeslee et
al.~\cite{Blakes10} have shown that a unimodal GC metallicity
distribution with an underlying mass-metallicity trend may create a
color magnitude relation with a distinct blue color peak --- a consequence
of the non-linear color-metallicity relation for GCs (see also
Richtler {\it et al.}~\cite{Richtl05}, Yoon {\it et al.}~\cite{Yoon06}, Peng {\it et al.}~\cite{Peng06}).

The most homogeneous and extensive sample of extragalactic GCs
currently available is that from the {\it ACS Virgo Cluster Survey}
(ACS VCS, C\^ot\'e {\it et al.}~\cite{Cote04}). From imaging of 100 Virgo
cluster early-type galaxies, more than 10000 GC candidates were
identified (e.g., Peng {\it et al.}~\cite{Peng06}, Jord\'an et
al.~\cite{Jordan09}). In Mieske {\it et al.}~\cite{Mieske06}, ACSVCS XIV, we examined
the color-magnitude relations of these GCs.  A highly significant
correlation, $\gamma_z \equiv {d(g-z)}/{dz} = -0.0374 \pm 0.004$ ($Z
\propto M^{0.48}$), between color and magnitude was found for the
subpopulation of blue GCs in the co-added samples of the three
brightest Virgo cluster galaxies (M49, M87 and M60). In general, the
trend was found to be more pronounced for GCs belonging to more
massive host galaxies, but there were also clear galaxy-to-galaxy
differences; significant blue tilts were found for the GC systems of
M87 and M60 but not for that of M49 (see also Strader et
al.~\cite{Strade06}). Peng {\it et al.}~\cite{Peng09} have recently
confirmed the presence of a blue tilt in M87 using much deeper (50
orbit) ACS imaging in the F606 ($V$) and F814 ($I$) filters, in
contradiction to a claim based on these same data that the blue tilt
is an observational artifact (Waters {\it et al.}~\cite{Waters09}).

In this paper, we analyze the color-magnitude relation of GCs in the
{\it ACS Fornax Cluster Survey} (ACS FCS, Jord\'an et
al.~\cite{Jordan07a}). This survey, which imaged 43 early-type
galaxies belonging to the Fornax cluster, is nearly identical in
design to the ACS VCS, although it targeted galaxies in a new and
different environment. We also analyze the combined GC samples from
the Fornax and Virgo surveys, to improve the overall statistics and
highlight possible cluster-to-cluster differences. In what follows, we
focus on four main topics: (1) the variation in the tilt slope as a
function of environment; (2) the mass scale where the tilt first
appears; (3) the constraints imposed on GC self-enrichment models
posed by our findings; and (4) the amount of tilt that may arise from
GC dynamical evolution.

\section{Selection of Globular Clusters from the ACS Fornax Cluster Survey}

The ACS FCS sample consists of 43 early-type (E, S0, dE, dE,N, dS0)
members of the Fornax cluster.  Each galaxy was imaged in the F475W
and F850LP filters ($\approx g_{475}$ and $z_{850}$) for a total of
760 and 1220 sec, respectively. This filter combination gives roughly
a factor-of-two improvement in wavelength baseline and metallicity
sensitivity compared to the ``canonical'' $(V-I)$ color index
(C\^ot\'e \etal\ 2004). The identification of bona-fide GCs from these
images is performed in the size-magnitude plane as described in Peng
\etal\cite{Peng06} and Jord\'an \etal\cite{Jordan09}. This selection
procedure --- which is possible because the half-light radii of GCs
are marginally resolved (Jord\'an \etal\cite{Jordan05a}) at the
distance of Fornax ($d =20.0$ Mpc; Blakeslee \etal\cite{Blakes09}) ---
greatly reduces contamination from both foreground stars and
background galaxies.

The ACS images have been reduced using a dedicated pipeline that is described 
in Jord\'an \etal (\cite{Jordan04a}, \cite{Jordan04b} and \cite{Jordan07a}.
In brief, the reductions consist of image combination, galaxy modeling,
model subtraction, rejection of obvious background galaxies, and the
measurement of magnitudes and sizes for candidate GCs using the
program KINGPHOT (Jord\'an \etal~\cite{Jordan05a}).
The result is a catalog containing integrated $g$ and $z$ magnitudes, $(g-z)$
colors and half-light radii, $r_h$, for all candidate GCs. Magnitudes
and colors are corrected for foreground extinction using the reddening
maps of Schlegel \etal~\cite{Schleg98} and extinction ratios for a G2
star as specified in Sirianni et al~\cite{Sirian05}.

To estimate the contamination by background galaxies, an identical
reduction procedure was applied to 17 blank, high-latitude fields
observed by ACS in the F475W and F850LP filters (see Peng
\etal~\cite{Peng06} and Jord\'an \etal ~\cite{Jordan09} for details).
The relative distributions of contaminants and GCs in the
size-magnitude plane then make it possible to assign a GC probability,
$\cal P_{\rm gc}$, to every source detected in each field.  In our
analysis, all sources with ${\cal P}_{\rm gc} \ge 0.5$ are considered
to be GCs.  The residual contamination from background galaxies that
are assigned ${\cal P}_{\rm gc} \ge 0.5$ is negligible for this study
(see Mieske {\it et al.}~\cite{Mieske06}, ACSVCS XIV in the following).

\section{Results}

Fig.~\ref{CMDfornax} shows color-magnitude diagrams (CMDs) for GCs in
the Fornax cluster in $M_z$ vs. $(g-z)$: for NGC 1399 alone, the
entire FCS sample, and the FCS sample separated by host galaxy
magnitude at $M_B=-19.75$ mag. This magnitude corresponds to a stellar
mass of $M_* \approx 5\times10^{10}M_{\odot}$ (Peng \etal\
2008). Although somewhat arbitrary, this choice divides the GC samples
into two nearly equal halves, while at the same time corresponding
roughly to several noteworthy mass scales in the physics of galaxy
formation: i.e., the approximate transition between ``young and old"
galaxies, at $M_* \sim 3\times10^{10}M_{\odot}$ (e.g., Kauffmann
\etal\ 2003, 2004), and the division between galaxies showing central
luminosity deficits and excesses in their surface brightness profiles,
at $M_* \sim $ (5-10)$\times10^{10}M_{\odot}$ (Ferrarese \etal\
2006ab; C\^ot\'e \etal\ 2006, 2007).  Throughout this paper, we will
refer to the galaxy sample with $M_B < -19.75$ mag as the {\tt
high-mass} sample, and that with $M_B \ge -19.75$ as {\tt low-mass}
sample.

We use the SBF distance moduli from Mei {\it et al.}~\cite{Mei07} and Blakeslee
\etal~\cite{Blakes09}, which have an internal precision of $\approx
0.08$~mag, to assign absolute magnitudes to the associated GCs.
Fig.~\ref{CMDfv} shows the CMDs of the full VCS sample,
the combined FCS and VCS sample, and the combined FCS and VCS sample after separation into
{\tt high-mass} and {\tt low-mass}  galaxies.

\subsection{\kmm Fitting Procedure}

\begin{figure*}
\centerline{\plottwo{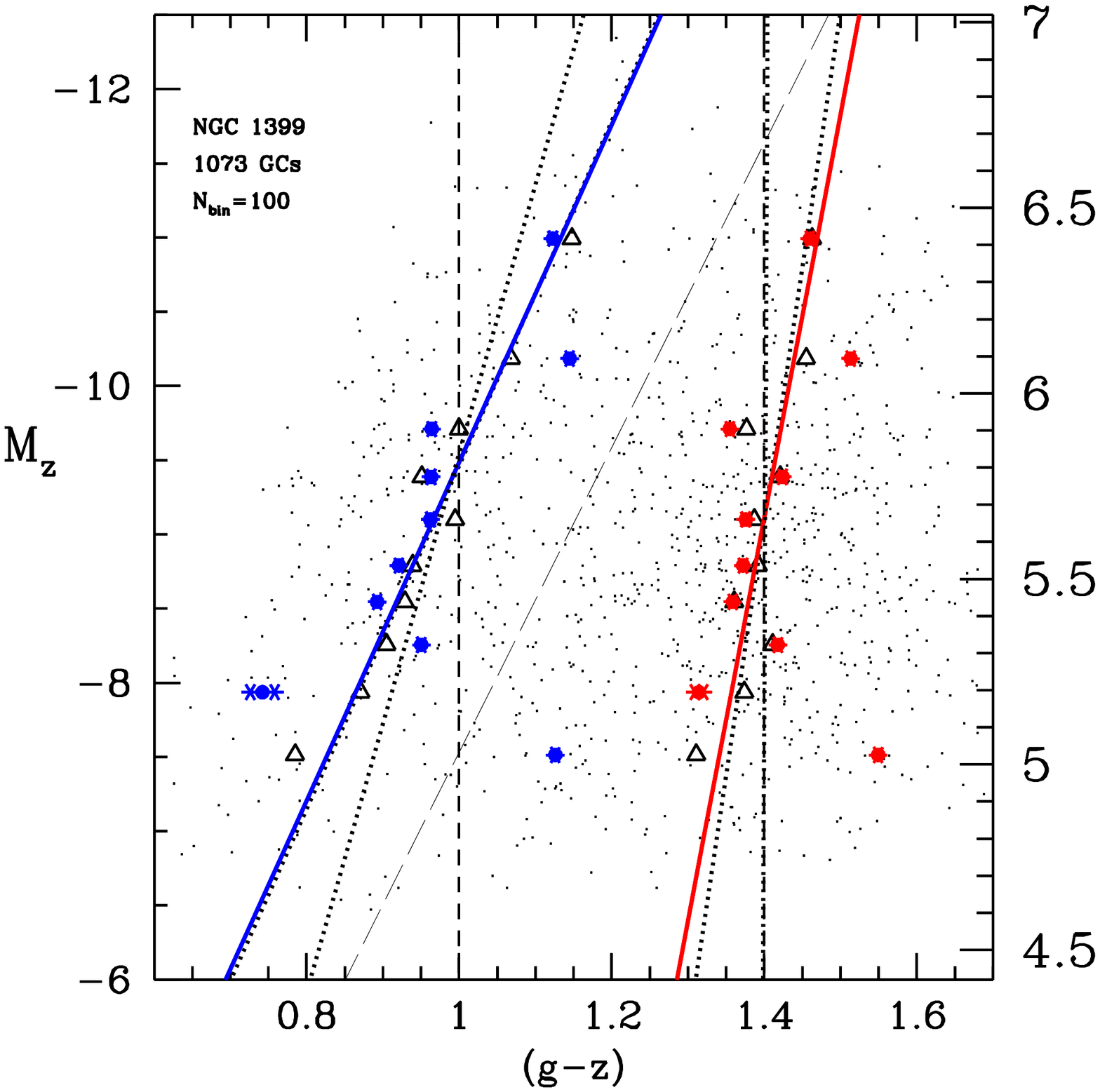}{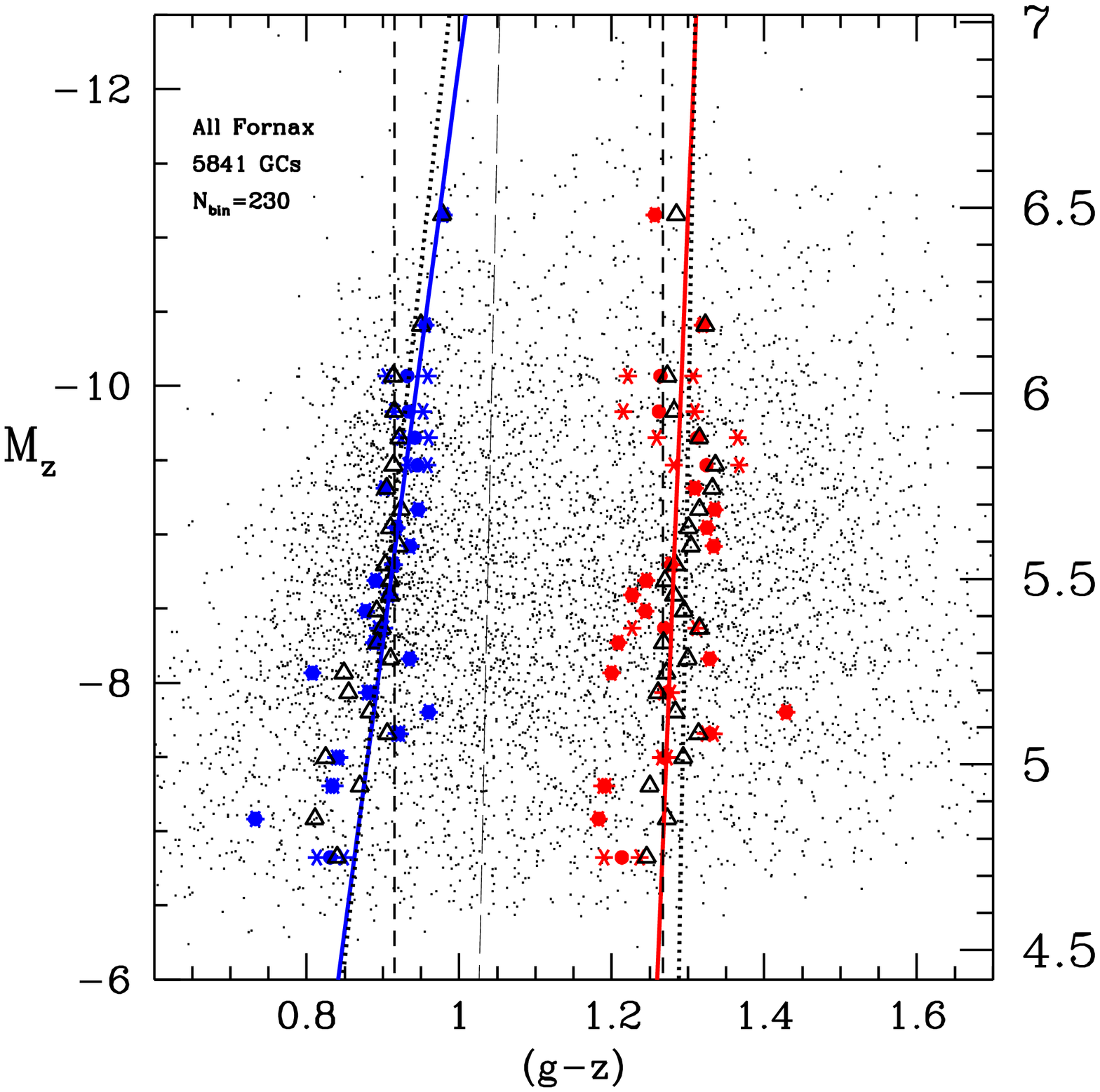}}
\centerline{\plottwo{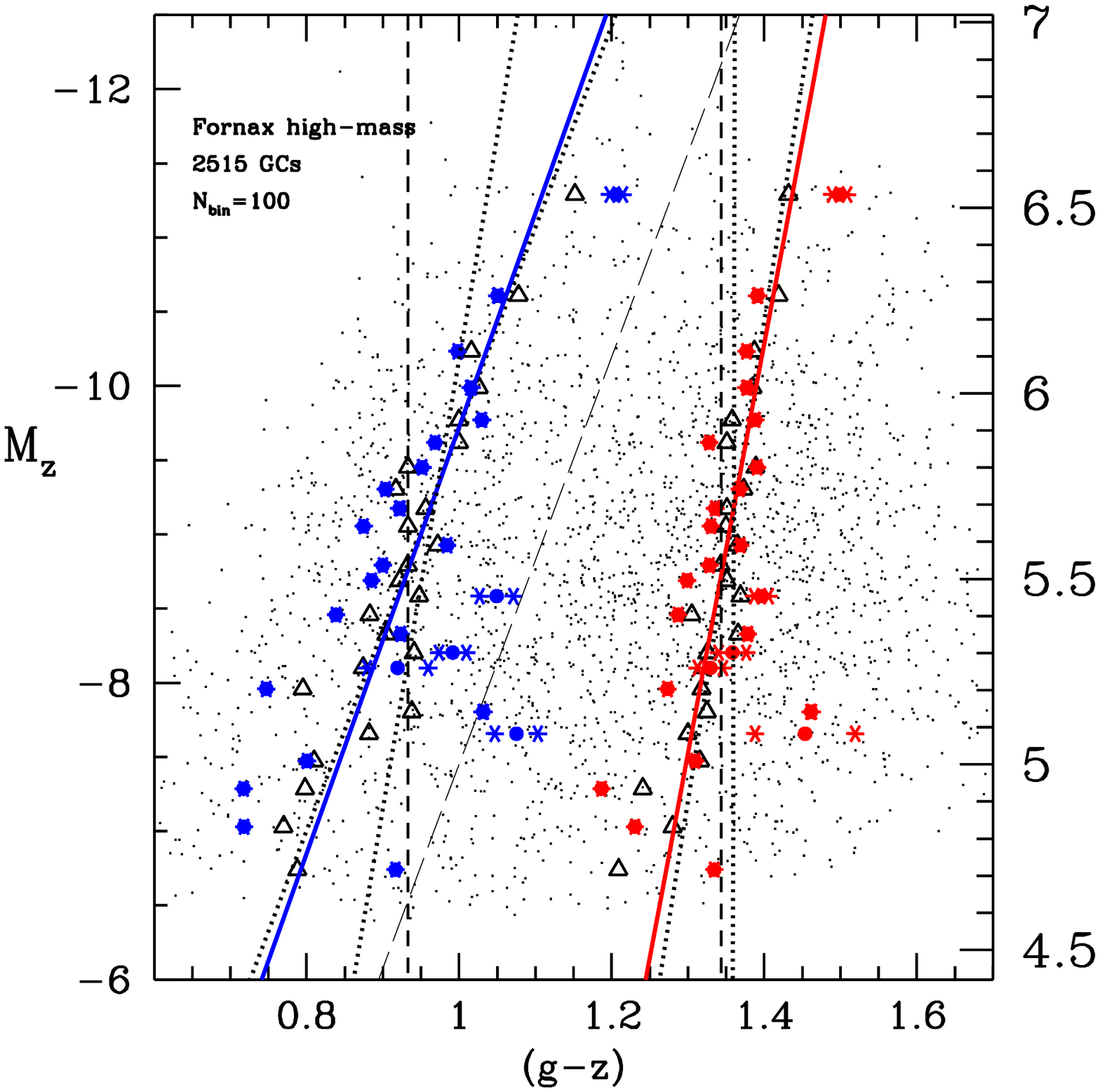}{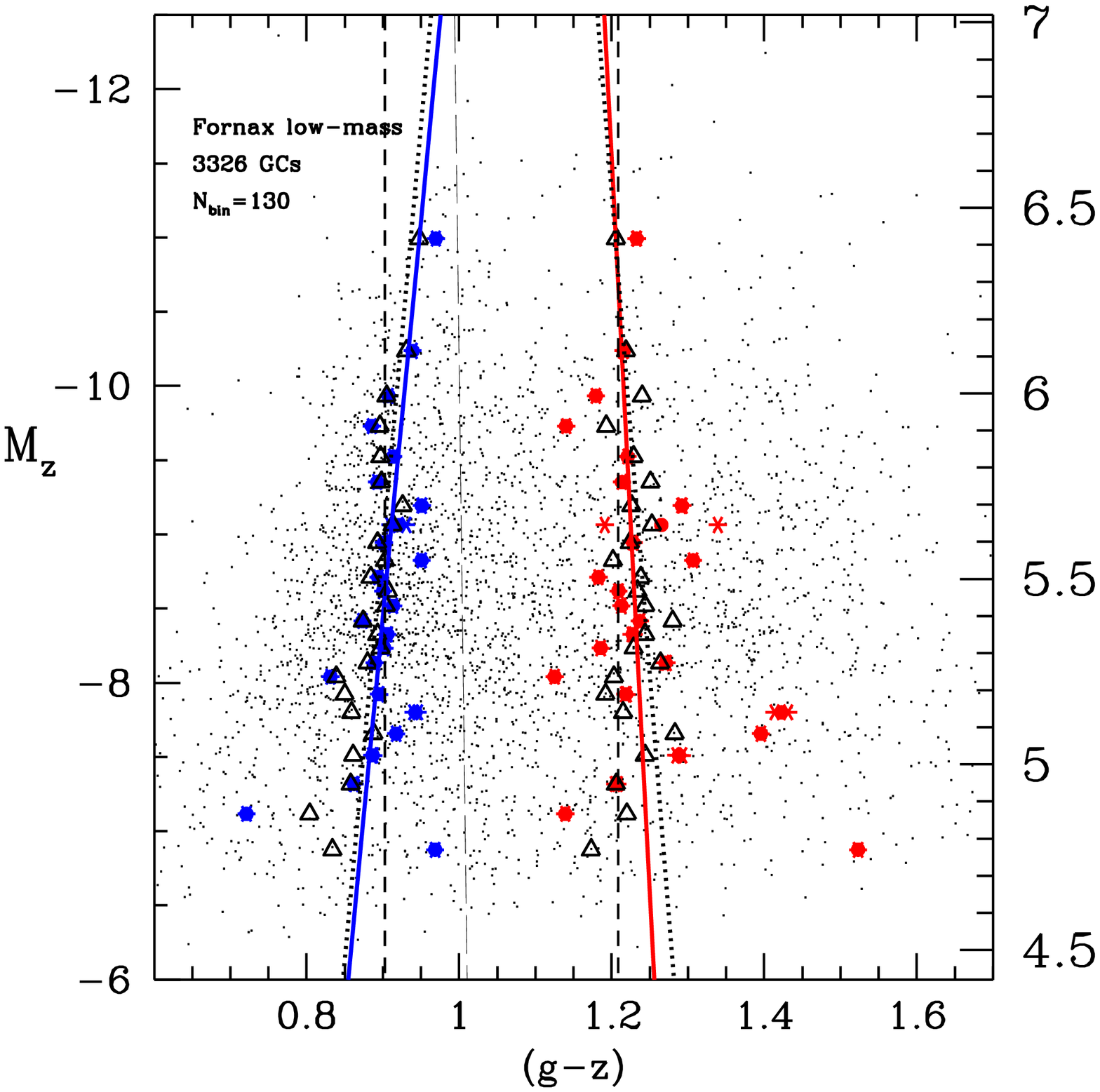}}
\caption{\label{CMDfornax}  $M_z$ vs. $(g-z)$ color-magnitude diagram (CMD)
for GCs detected in the ACS Fornax Cluster Survey (FCS). From {\it top left}
to {\it bottom right}, the plots show: (1) the GCs associated with NGC 1399; (2) all GCs in the
FCS; (3) GCs belonging to {\tt high-mass} galaxies; and (4) GCs belonging to
{\tt low-mass} galaxies. The vertical dashed lines in each panel mark the
color expected for the blue and red populations based on the
relation between host galaxy magnitude and peak colors presented in
Peng {\it et al.}~\cite{Peng06}.  Filled circles indicate the \kmm fitting
results, corresponding to the means obtained from adopting two
different initial guesses (indicated by asterisks; see \S3.1
for more details). The number of luminosity bins is 25, except for NGC
1399, where 10 luminosity bins were adopted. The solid blue and red
lines indicate linear least-squares fits to the \kmm peak
positions. The long dashed line at intermediate colors denotes the
dividing point between the blue and red peak (i.e. the color where
both blue and red peak contribute 50\% to the number counts) 
derived from the \kmm fits. The open triangles indicate the median
color of GCs blueward and redward of the dividing line. The dotted line
is a fit to the median colors. For both  the CMD of NGC~1399 and that of the FCS
{\tt high-mass} sample, we also show least-squares fits to the median
colors blueward and redward of a magnitude independent dividing line. That dividing line is given
by the mean color of the long dashed line between the brightest magnitude bin and $M_z=-8.1$ mag.}
\end{figure*}

Analogous to ACSVCS XIV, we apply \kmm (Ashman {\it et
al.}~\cite{Ashman94}) in heteroscedastic mode to the CMDs of these
sub-samples, subdivided into 25 luminosity bins containing the same
number of GCs. By adopting a fixed number of luminosity bins, we are
able to get a straightforward comparison of the color scatter between
different subsamples. The only exception is the CMD of NGC 1399 in
Fig.~\ref{CMDfornax}, for which we adopt 10 luminosity bins due to the
smaller number of GCs.  For each bin, \kmm is used to determine the
best-fit blue and red peak magnitudes of a bimodal Gaussian color
distribution. In the heteroscedastic mode, we allow for Gaussians of
different widths to be fitted to the blue and red GC subpopulations.

\begin{figure*}
\centerline{\plottwo{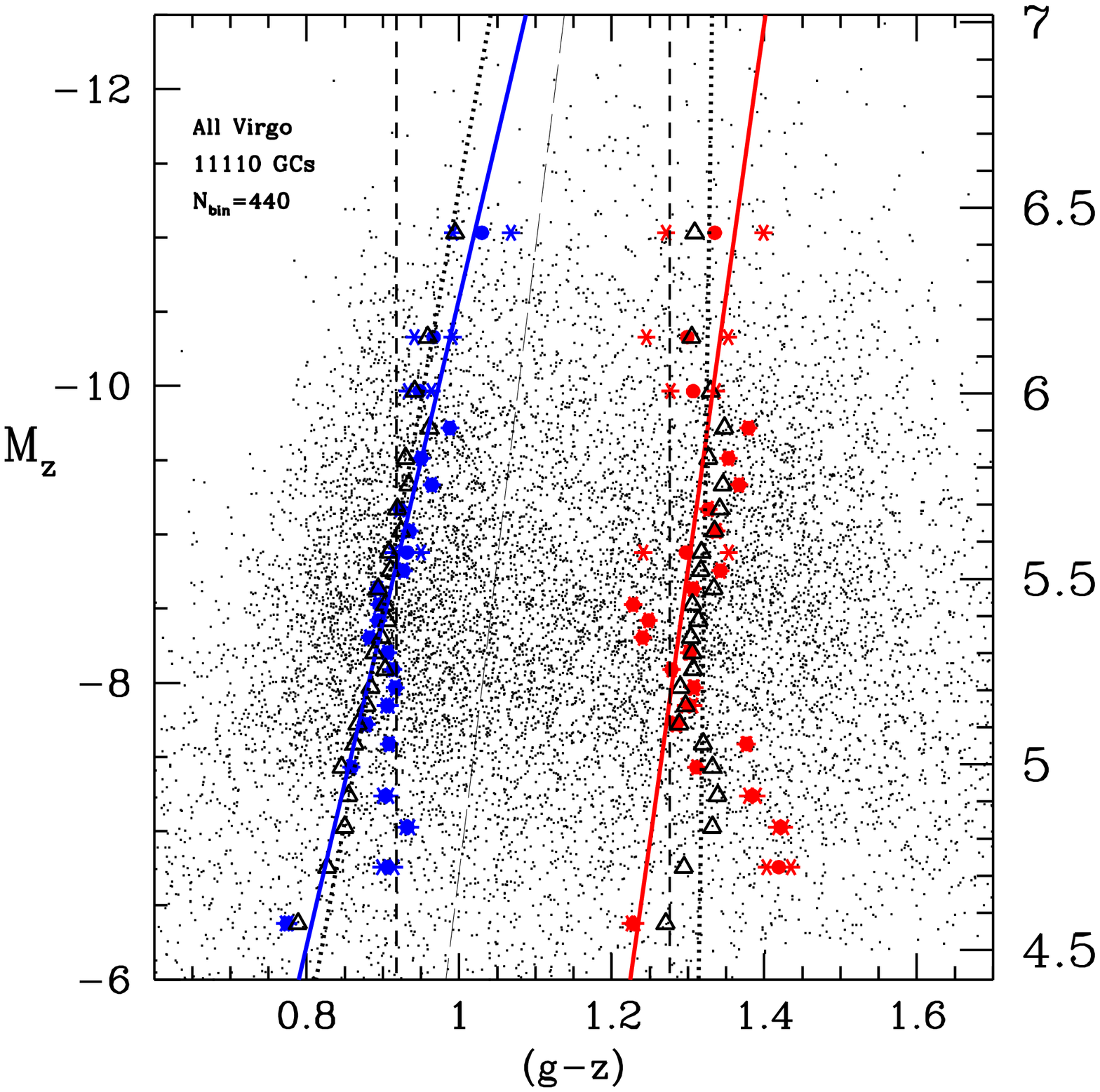}{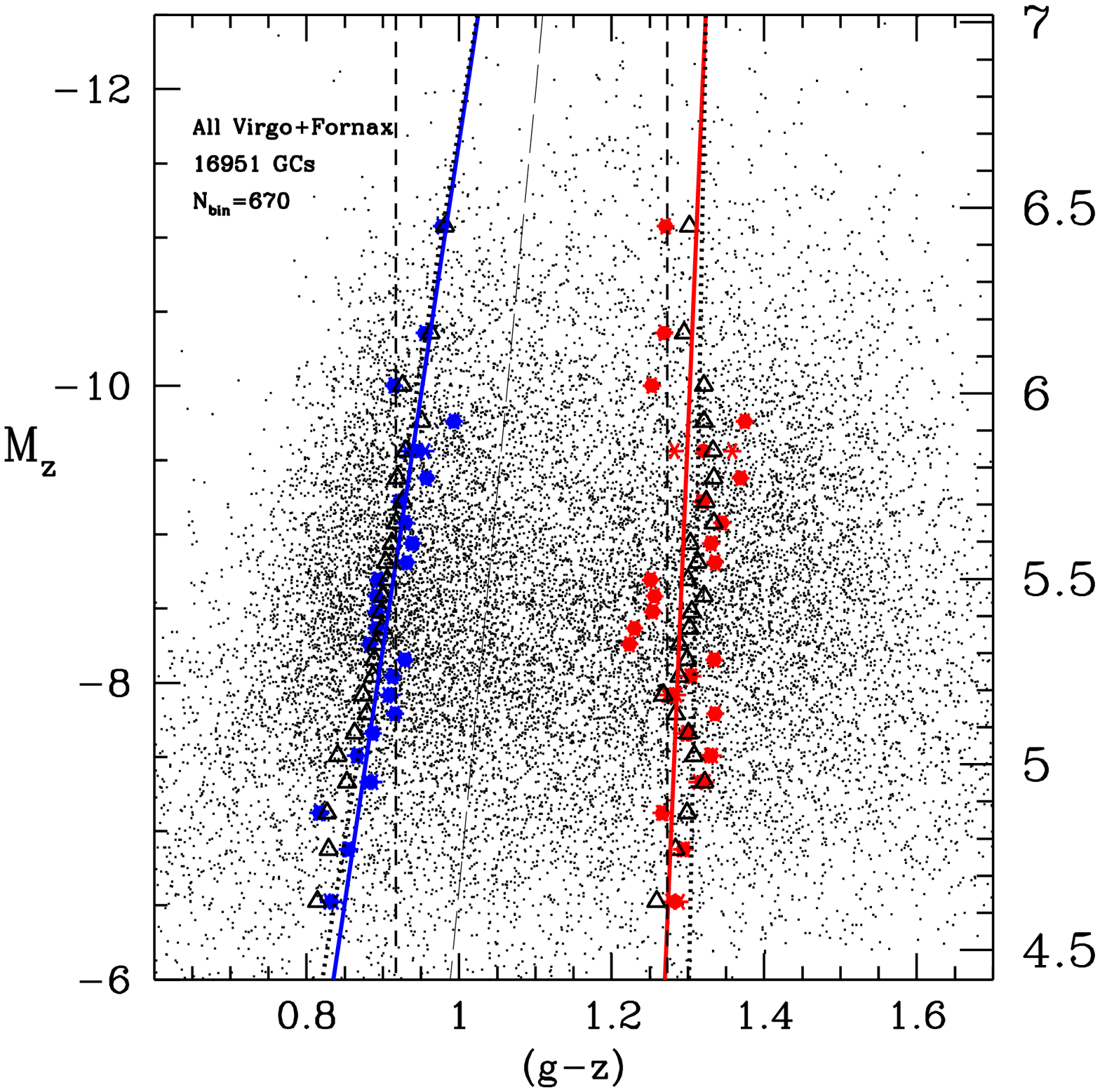}}
\centerline{\plottwo{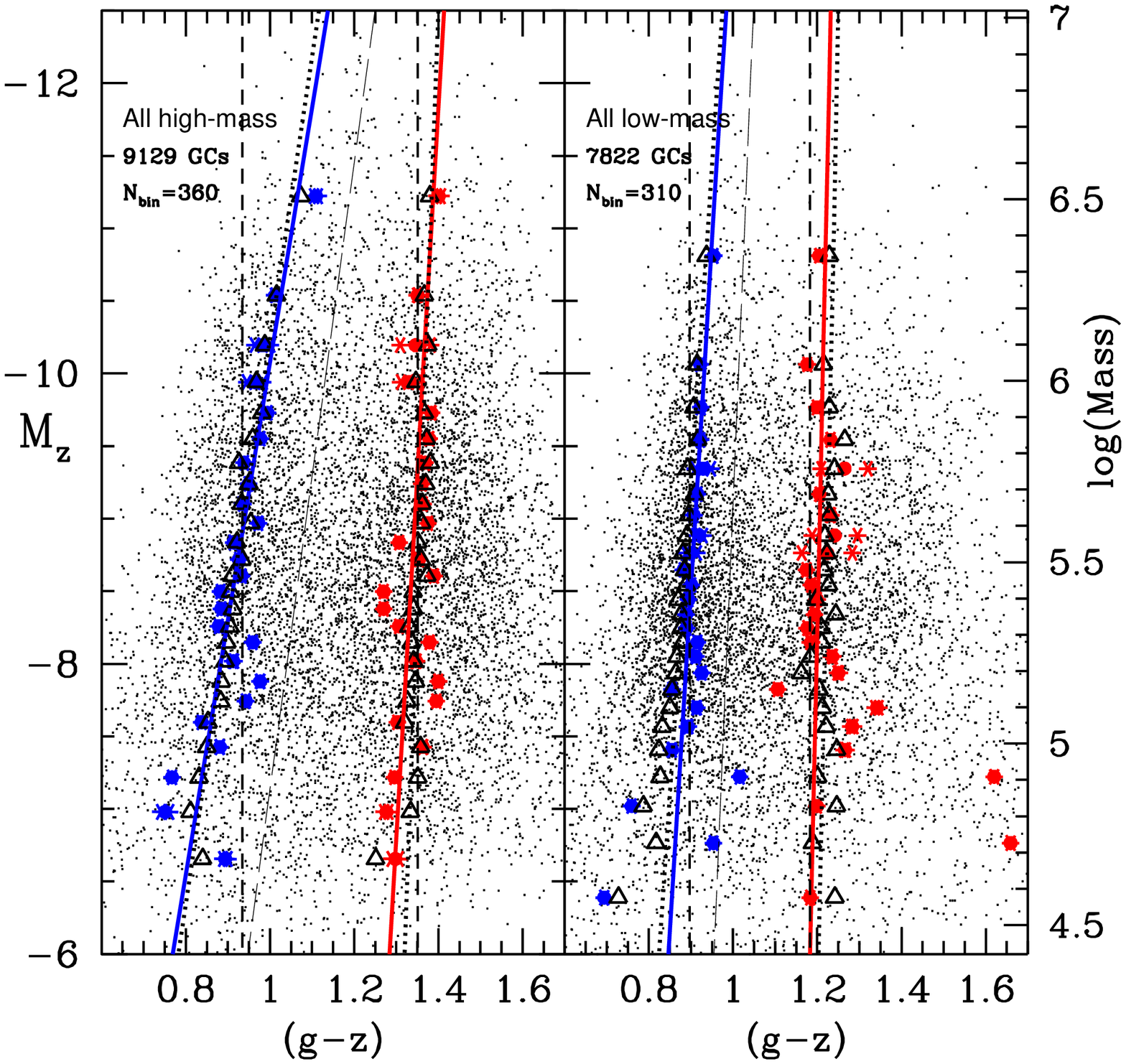}{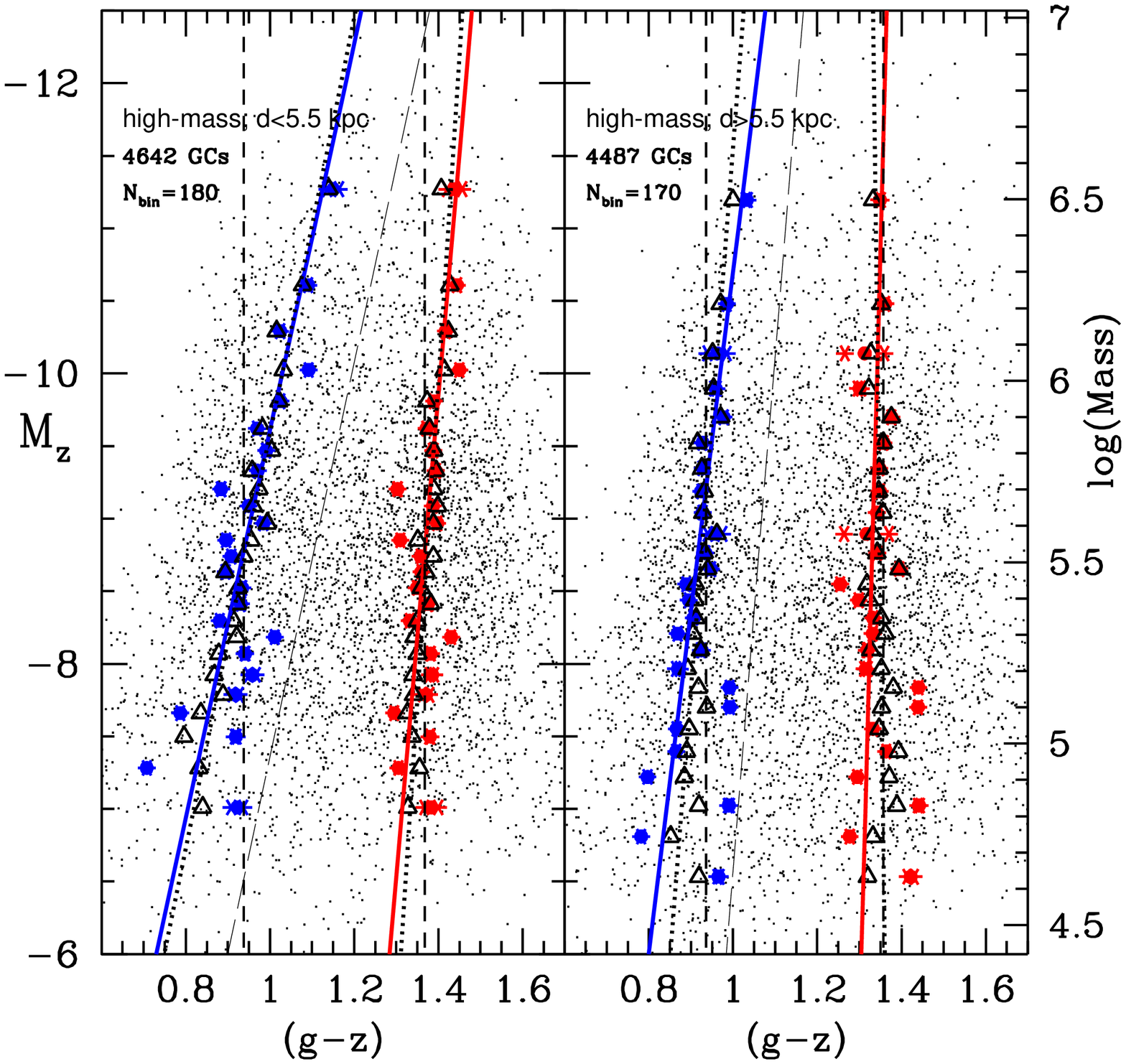}}
\caption{\label{CMDfv} CMDs for various GC subsamples with results from
\kmm overplotted. The symbols and lines are the same as in Fig.~\ref{CMDfornax}. 
{\it (Upper Left Panel).} The full VCS sample.
{\it (Upper Right Panel).} The combined FCS and VCS sample.
{\it (Lower Left Panel).} The left portion of the diagram shows all GCs from the {\tt high-mass}
FCS and VCS galaxies, while the right portion shows GCs
from the {\tt low-mass} FCS and VCS galaxies. Note the
more pronounced tilt for GCs belonging to the {\tt high-mass} galaxies. 
{\it (Lower Right Panel):}
CMD showing two subsets of GCs in the {\tt high-mass} sample, divided
at a projected galactocentric distance of $d=$ 5.5kpc into inner and
outer samples.}
\end{figure*}

To obtain an estimate of the uncertainty involved in the choice of the
initial peak position guesses, we perform the \kmm fitting in two
iterations. We first run \kmm using the mean peak colors of the full
samples (i.e., FCS, VCS, or FCS and VCS combined) as initial
guesses. From the KMM results for a given subsample, we then calculate
the peak color $(g-z)_0$ averaged over all luminosity bins, and
calculate the dispersion, $\sigma_0$, of the peak colors around their
mean. Typically, $\sigma_0$ was found to be in the range 0.05 to 0.1
mag. For the second iteration, we then run \kmm twice, with two
different initial guesses: $(g-z)_0 \pm \sigma_0$.  For the final
measurement we adopt the mean of the \kmm results obtained from the
two initial guesses. In Figs.~\ref{CMDfornax} and \ref{CMDfv} these
fitted positions of the blue and red peaks are plotted over the
respective CMDs. The mean is indicated by a filled circle, while the
results from the two initial guesses are shown as asterisks.  In most
cases, the \kmm results were found to be identical for the two initial
guesses.  We also note that for all the samples investigated in this
paper, {\it bimodal} Gaussian color distributions are strongly
prefered over {\it unimodal} Gaussian distributions (confidence levels
$\gtrsim$99$\%$). Note that in ACSVCS XIV, it was shown that the \kmm
results are reliable for $M_z$-$(g-z)$, while for $M_g$-$(g-z)$ they
are biased toward progressively bluer colors at bright luminosities
(due to the depopulation of the red peak at high $g$-band
luminosities). We therefore do not consider $M_g$-$(g-z)$ CMDs.

In these plots, we show as vertical dashed lines the mean color
expected for the blue and red peak based on the relation between host
galaxy magnitude and GC peak colors found by binning independent color
distributions (Peng \etal \cite{Peng06}). The mean colors of the blue
peaks agree very well with the \kmm fit results in this paper. For the
full VCS, and VCS plus FCS, samples, the \kmm fits and median colors
give a small redward offset of $\lesssim 0.05$ mag for the red peak
compared to the results of Peng {\it et al.}~\cite{Peng06}, due to the
slightly asymmetric shape of this peak (see Figs.~2 and 5 of Peng
\etal \cite{Peng06}).

To measure the amplitude of the tilt, we perform least-square fits of
the KMM data points for the blue and red peaks as a function of
magnitude, excluding GC candidates brighter than $M_z=-12.5$ mag and
magnitude bins fainter than $M_z=-8.1$~mag. The resulting linear
relations are indicated in the CMDs of Figs.~\ref{CMDfornax} and
\ref{CMDfv}. The errors on the fit are calculated by resampling the
points using as the dispersion the observed scatter around the fitted
relation. The measured slopes and errors $\gamma_z \equiv
{d(g-z)}/{dz}$ are shown in Table~\ref{Fitresults}. We also indicate
an estimate of the corresponding mass-metallicity scaling $Z \propto
M^n$. This is obtained from converting the $(g-z)$ datapoints to
[Fe/H] using a biquadratic transformation based on Peng
\etal~\cite{Peng06} and fitting a linear relation to the data points
in magnitude-[Fe/H] space. The slope of this transformation is
$\frac{\Delta \rm [Fe/H]}{\Delta \rm (g-z)}\simeq 5$ dex/mag in the
colour range of the blue peak ($(g-z)\simeq 0.95$ mag) and
$\frac{\Delta \rm [Fe/H]}{\Delta \rm (g-z)} \simeq 2$ dex/mag in the
colour range of the red peak ($(g-z)\simeq 1.35$ mag). We assume a
constant mass-to-light ratio over the luminosity range investigated.

As a check on the \kmm fits, we calculate the slope using the median
colors blueward and redward of the limiting color between the blue and
red peak. In Figs.~\ref{CMDfornax} and \ref{CMDfv}, the median values
are indicated as open triangles, and the fit to them as a dotted
line. Note that we define the limiting color as the point where the
contribution from the blue and red peak is equal; this is generally
different from the minimum of the color distribution. The limiting
color is first determined for each luminosity bin based on the results
from {\tt KMM}. We then fit a linear relation to the limiting color as
a function of magnitude and use this limiting color (the long dashed
lines in Figs.~\ref{CMDfornax} and \ref{CMDfv}) to separate the blue
and red peaks. We report the slopes of these median estimates in
Table~\ref{Fitresults}. These generally agree quite well with the \kmm
results. Only the median in the most luminous bin is somewhat bluer
than the \kmm result. This is because the \kmm fit in the brightest
bin typically shows a redward ``jump" of the color distribution with
respect to fainter magnitudes (see Sect.~\ref{onsetmass}) due to the
near merging of the blue and red peak, an effect that is not as
apparent using the median estimate due to the smoothly changing colour
limit adopted for the median calculation. The blue tilt measured from the
median is hence, on average, 8-10\% weaker than that measured from
{\tt KMM}.  For the CMDs of NGC 1399 and the Fornax {\tt high-mass}
sample --- which exhibit the most dramatic tilts --- we also indicate
the ``minimum'' color-magnitude relations obtained from determining
the median colors blueward and redward of a magnitude-independent
limiting color.

\subsection{Measured Slopes for Various Subsamples}
\label{subsamples}

Table~\ref{Fitresults} shows that the slope obtained for NGC~1399 is
quite dramatic, with $\gamma_z= -0.0878 \pm 0.0250$. This trend is
significant at roughly the 3.5$\sigma$ level. In mass-metallicity
space, this corresponds to a scaling relation of $Z \propto M^{0.82
\pm 0.23}$. The morphology of the CMD of NGC~1399 shows that this
strong slope originates from a near-merging of the blue and red peaks
at the highest luminosities (see also Dirsch \etal \cite{Dirsch03};
Bassino \etal \cite{Bassin06}).  Nevertheless, we emphasize that \kmm
still gives a confidence level of $98-99\%$ for prefering a bimodal
Gaussian over a unimodal Gaussian for the brightest bins.

When considering the full sample of GCs from the FCS, we find a
shallower but still highly significant tilt of $\gamma_z=-0.0257 \pm
0.0050$. This is weaker than the tilt $\gamma_z=-0.0459 \pm 0.0048$
found for the full VCS data set (see Fig.~\ref{CMDfv} and
Table~\ref{Fitresults}).  For the subsample of GCs belonging to the
{\tt high-mass} FCS galaxies, we find a slope of $\gamma_z=-0.0697 \pm
0.0116$.  GCs belonging to the {\tt low-mass} FCS galaxies show a
rather weaker slope of $\gamma_z=-0.0186 \pm 0.0066$.  This result for
the FCS confirms the finding from ACSVCS XIV that the slope
becomes weaker for GCs belonging to less luminous galaxies. This also
holds when considering the alternative fit results based on median
colors, and even when using the ``minimum'' slope derived in the
previous subsection. To reinforce this point, the bottom left panel of
Fig.~\ref{CMDfv} shows the combined FCS and VCS sample, subdivided
into {\tt high-mass} and {\tt low-mass} galaxies. The tilt for the
{\tt high-mass} sample is $\gamma_z=-0.0566 \pm 0.0076$, almost three
times steeper than that for the {\tt low-mass} sample,
$\gamma_z=-0.0210 \pm 0.0034$. The two slopes are inconsistent with
each other at the 4.3$\sigma$ level, and imply a range of mass
metallicity scalings $Z \propto M^{0.3-0.7}$ as a function of galaxy
luminosity/mass.  For the full sample of GCs from the two surveys
(Fig.~\ref{CMDfv}), we find a slope of $\gamma_z=-0.0293 \pm
0.0085$. This slope, which is, as expected, intermediate to those
found for the FCS and VCS samples, corresponds to $Z \propto M^{0.43
\pm 0.12}$.

The bottom right panel in Fig.~\ref{CMDfv} shows two GC subsets in the
{\tt high-mass} galaxies from the combined FCS and VCS samples,
separated at a projected galactocentric distance $d=5.5$ kpc
(corresponding to $\approx$ 0.25--1$R_e$ for these galaxies; Ferrarese
\etal\ 2006a, 2010).  The slope found for the {\it inner sample} is
$\gamma_z=-0.0750 \pm 0.0120$, significantly higher than the value
$\gamma_z=-0.0428 \pm 0.0071$ found for the {\it outer sample}. The
tilt difference is even more pronounced when using the alternative
median fitting. This confirms the findings from ACSVCS XIV
that the tilt is most significant for GCs lying at small projected
galactocentric distances.

\subsubsection{M49 Revisited: Blue Tilt Differences Between High-Luminosity Galaxies}

\begin{figure}
\centerline{\plotone{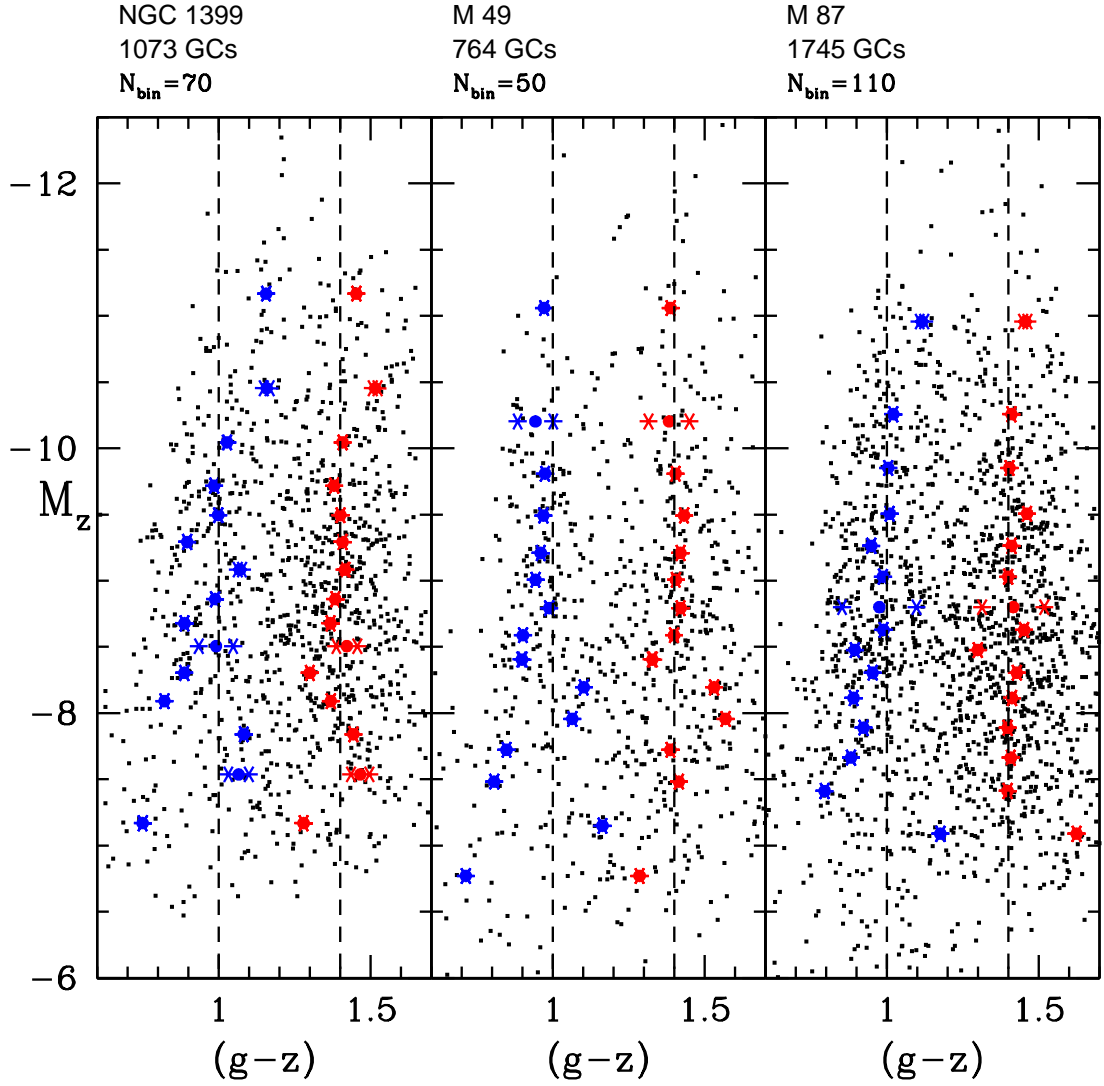}}
\caption{\label{3giants} Distribution of GCs in the CMD for three giant elliptical galaxies
NGC~1399 (Fornax), M49 (Virgo), and M87 (Virgo). Results from \kmm are
overplotted for a fixed number of 15 luminosity bins. Note the absence
of a blue tilt in M49, possibly due to the lack of intermediate-color GCs at
luminosities of $M_z \lesssim -10.5$ mag.}
\end{figure}

Both in ACSVCS XIV and in Strader {\it et al.}~\cite{Strade06} it was noted that
no blue tilt was detected for the brightest Virgo cluster galaxy
M49. To put this intriguing result into perspective, we compare in
Fig.~\ref{3giants} the CMDs for NGC 1399 (Fornax), M49 (Virgo), and
M87 (Virgo).  We choose a subdivision into 15 bins for the KMM fitting
to all three galaxies.  Both NGC~1399 and M87 show significant blue
tilts (ACSVCS XIV, Strader {\it et al.}~\cite{Strade06}, and Peng et
al.~\cite{Peng09}), while M49 apparently shows no obvious
color-magnitude relation. Comparing the overall morphology of the
CMDs, it is clear that in the case of M49, the lack of bright,
intermediate-color GCs is at least partly responsible for the lack of
a tilt: i.e., the blue peak in M49 is narrow and there is a clear gap
between the blue and red peaks over the entire magnitude range.

This difference relative to M87 and NGC~1399 is difficult to reconcile
within a scenario where a GC mass-metallicity relation is created
entirely by a single mechanism operating within individual GCs (e.g.,
self-enrichment; see below).  In all likelihood, the detailed
accretion and merger histories of individual galaxies will also play a
role in establishing the distribution of GCs within the
color-magnitude plane.

\subsection{A Luminosity/Mass Threshold for the Blue Tilt} 
\label{onsetmass}

\begin{figure*}
\centerline{\plottwo{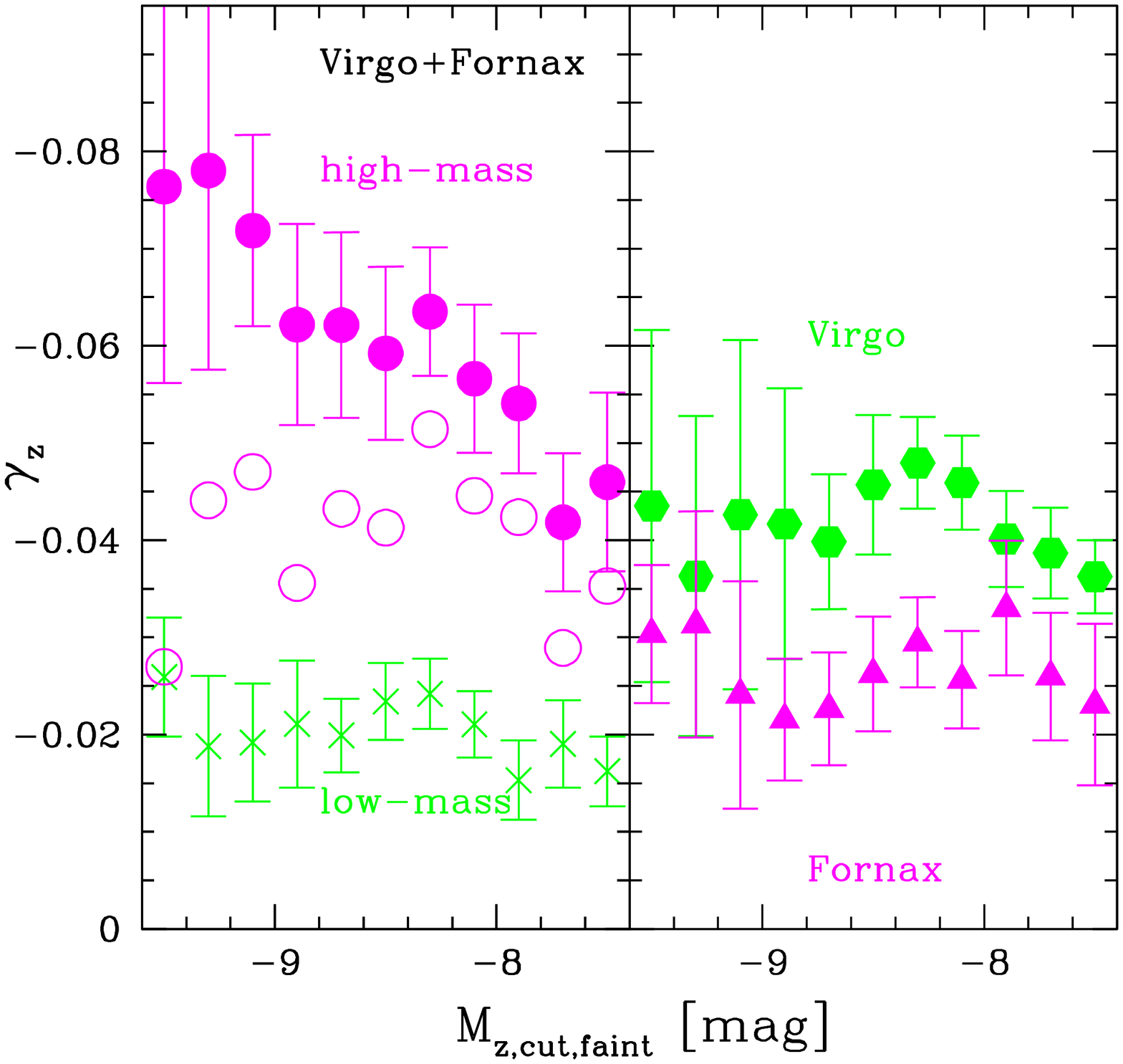}{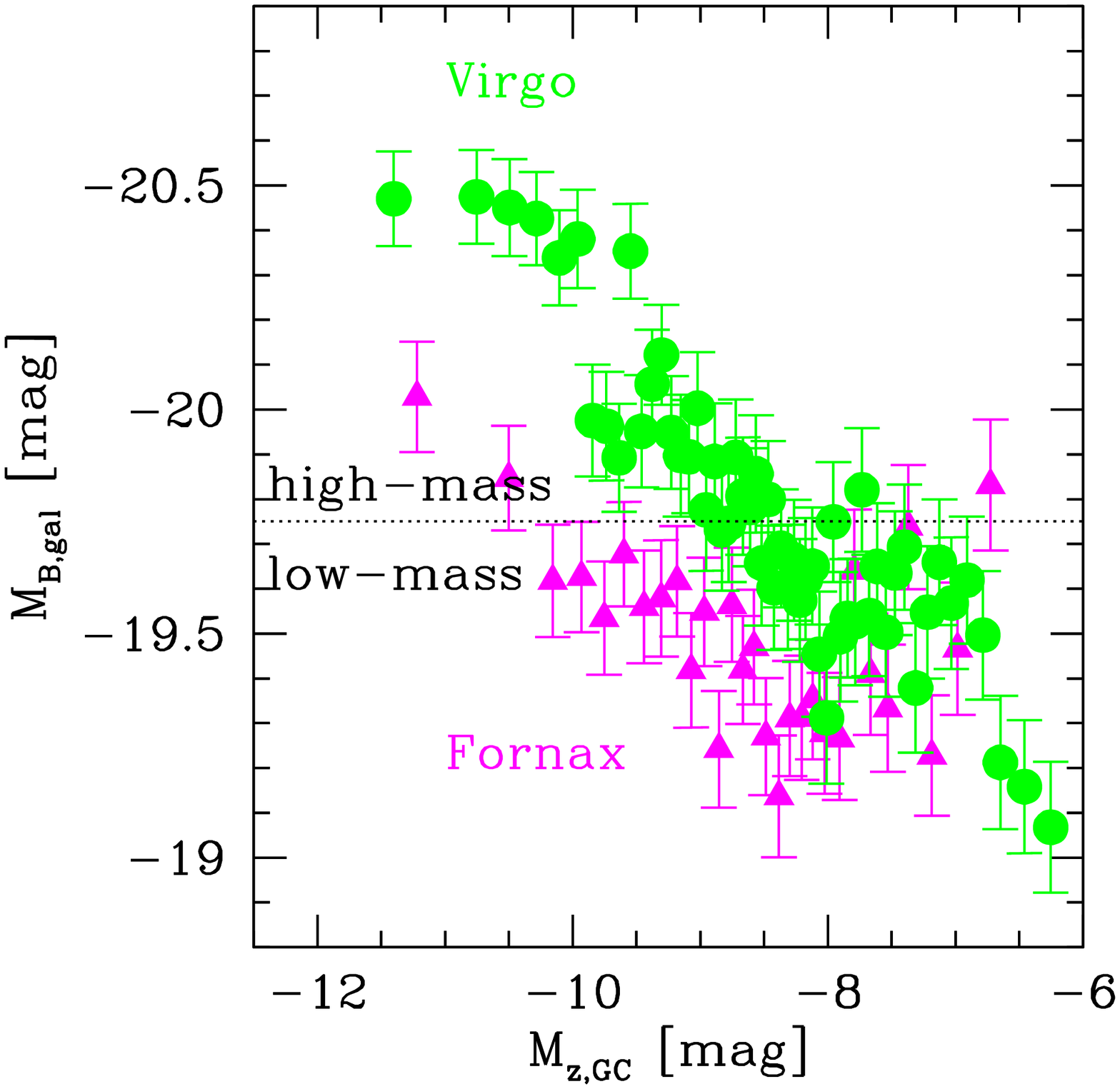}}
\centerline{\plottwo{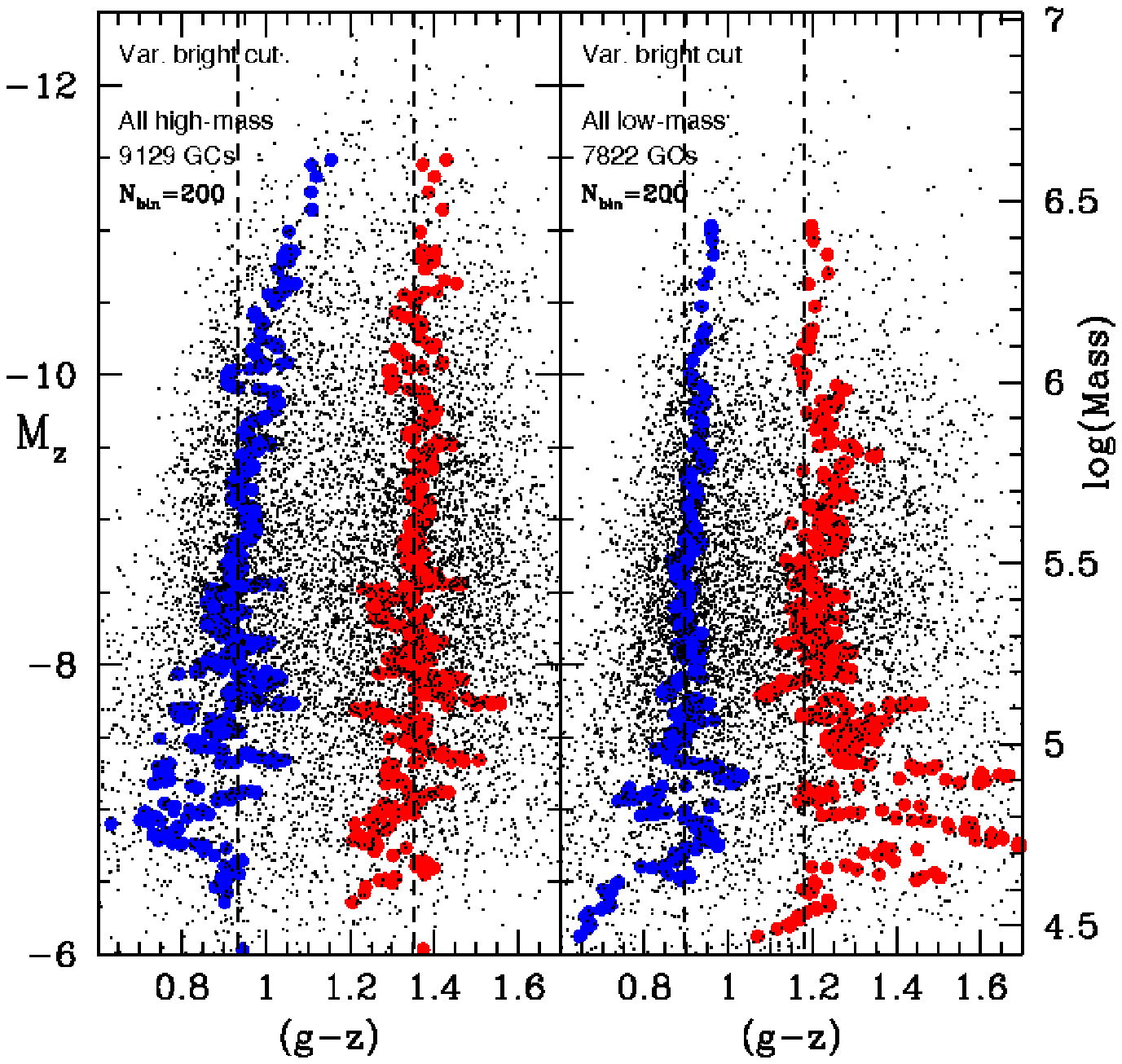}{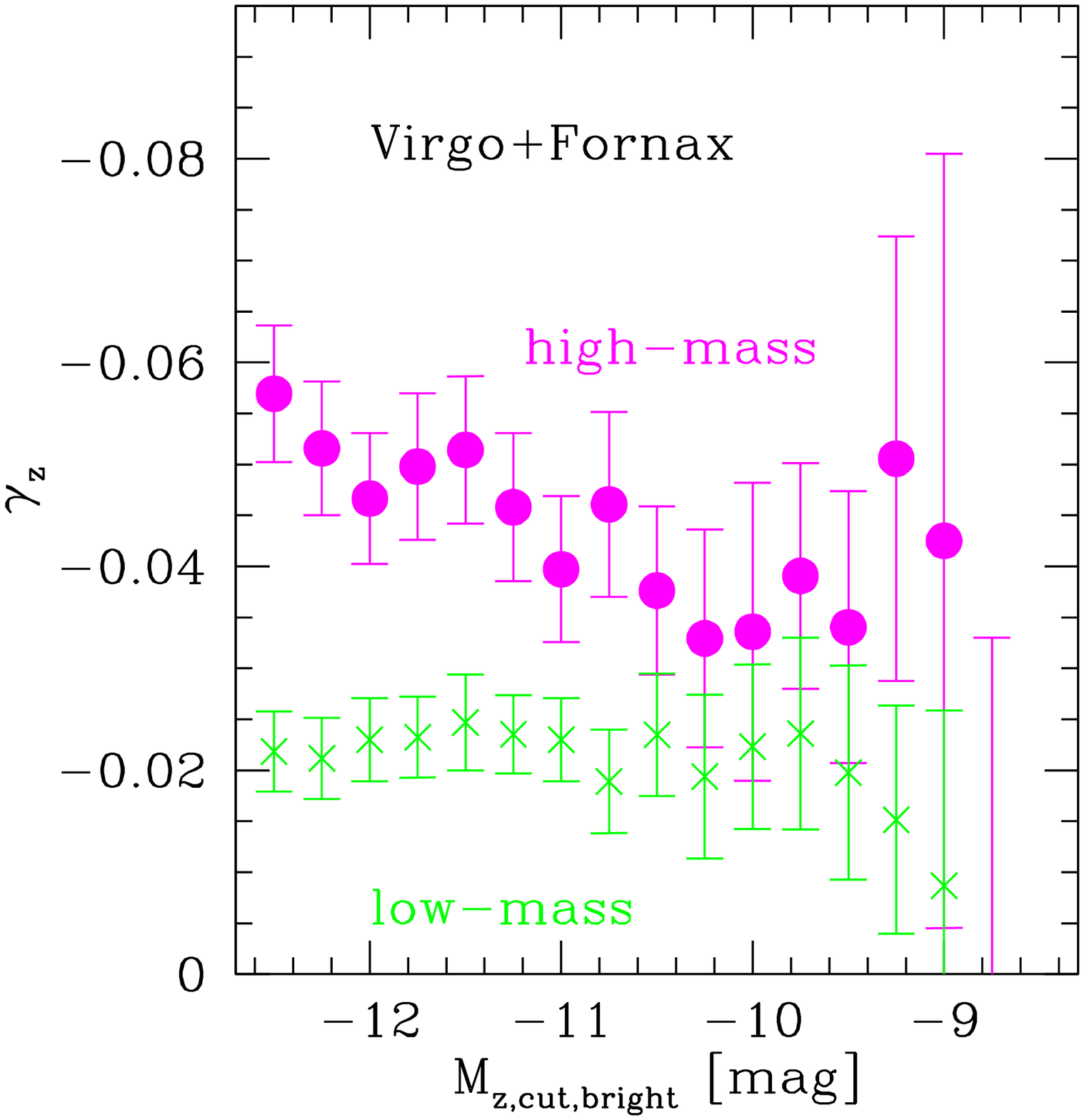}}

\caption{\label{slopes}Dependence of blue tilt on GC mass range used
in the analysis.  {\it (Upper Left Panel).} Variation in blue tilt
slope, $\gamma_z$, as a function of the faint magnitude cutoff,
$M_{\rm z,cut,faint}$. In the left portion of the diagram, the filled
magenta circles (green crosses) indicate the combined FCS and VCS
sample restricted to GCs belonging to {\tt high-mass} ({\tt low mass})
galaxies. Open circles indicate a fit to the {\tt high-mass} sample
when disregarding the brightest GC magnitude bin at $M_z=-11.2$ mag
(see the bottom left panel of Fig.~\ref{CMDfv}). In the right portion
of the diagram, the filled green circles show the full VCS sample,
while the filled magenta triangles show the full FCS sample.  {\it
(Upper Right Panel).} The mean host galaxy magnitude for GCs is
plotted vs. GC magnitude, for bin size of 200 GCs.  Filled green
circles indicate the full VCS sample, while filled magenta triangles
indicate the full FCS sample. The limiting magnitude between {\tt
high-mass} and {\tt low-mass} galaxies is indicated by a horizontal
dotted line.  {\it (Bottom Left Panel).} The CMDs of GCs belonging to
{\tt high-mass} and {\tt low-mass} galaxies. This is identical to the
bottom left panel of Fig.~\ref{CMDfv}, but with an overplotted
complementary set of \kmm fit results, in which we have varied the
bright magnitude limit of the GC samples between $M_z=-12.5$ mag and
$M_z=-8.5$ mag in steps of 0.25 mag. For each restricted sample, we
run \kmm and plot the fitted peak results.  {\it (Bottom Right
Panel).} This plot is based on the fits in the bottom left panel. It
shows the slope $\gamma_z$ fitted to the \kmm peak positions as a
function of the {\it bright} magnitude cutoff, $M_{\rm z,cut,bright}$,
applied to the GC samples. The faint limiting magnitude used in the
fitting is $M_z=-8.1$ mag. }

\end{figure*}

It has been suggested that the blue tilt is driven mainly by GCs that
are brighter than the GCLF turnover magnitude, (e.g. Harris et
al.~\cite{Harris06}, ACSVCS XIV, Peng et
al.~\cite{Peng09}). To investigate the luminosity/mass threshold of
the blue tilt, we show in the upper left panel of Fig.~\ref{slopes}
the dependence of slope $\gamma_z$ on the lower magnitude limit used
in the fitting procedure. In doing so, we have fitted a linear
relation to the \kmm peak results for the combined Fornax and Virgo
samples, subdivided into {\tt high-mass} and {\tt low-mass} host
galaxies (bottom left panel of Fig.~\ref{CMDfv}). The faint magnitude
cutoff has been varied from $M_{z,cut,faint}=-7.5$ mag to $-9.5$ mag.

For the {\tt high-mass} sample, there is a notable trend for the
derived slope to {\it increase steadily from $\gamma_z \simeq -0.045$
to $-0.075$ as the faint cutoff magnitude is varied from
$M_{z,cut,faint}=-7.5$ mag to $-9.5$ mag}. This may be a signature of
a non-linear color-magnitude relation at the upper end of the GC
luminosity function. The open circles in this plot show the change in
slope when the brightest magnitude bin, at $M_z=-11.2$ mag, is
excluded. In this case, the slope is roughly constant at $\gamma_z
\simeq -0.045$, a value of that is largely independent of the faint
cutoff for $M_{z,cut,faint} \lesssim -8.0$.  Hence, the steepening
slope seems to be due to the increasing weight of the very brightest
GCs, with $M_z \lesssim -11$ mag.

We conclude from this exercise that the blue tilt is firmly in place
above $M_{z,cut} \approx -$8 mag ($1.5-2 \times 10^5$M$_{\sun}$) and
further steepens at $M_z \lesssim -11$ mag ($\gtrsim 2-3 \times
10^6$M$_{\sun}$).  Nevertheless, the upper left panel of
Fig.~\ref{slopes} shows clearly that the strong blue tilt for GCs in
the {\tt high-mass} sample is not driven by the brightest GCs alone
since  the open circles are consistently above the crosses.

This plot also demonstrates once again that the blue tilt is more
pronounced in the Virgo data. In this context it is crucial to note
that the GCs in the VCS sample originate, on average, from more
luminous galaxies than in the FCS sample, see the upper right panel of
Fig.~\ref{slopes}: there is a difference of 0.5-1.0 mag in host galaxy
magnitude. The GC number ratios between the {\tt high-mass} and {\tt
low-mass} samples is 43:57 for Fornax, but 60:40 for Virgo. Given the
overall stronger tilt found for the {\tt high-mass} samples, this
difference in host galaxy distribution appears to be the root cause of
the stronger tilt found using the Virgo sample.

In the bottom panels of Fig.~\ref{slopes} we investigate in detail how
$\gamma_z$ changes as a function of the {\it bright} cutoff
magnitude. We consider the {\tt high-mass} and {\tt low-mass}
subsamples in the combined FCS and VCS sample and keep $M_z=-8.1$ mag
fixed as the faint cutoff. In order to obtain a finer grid of peak
positions at bright luminosities we: (1) apply a varying bright cutoff
to the individual GC data points, instead of a cut to the \kmm peak
positions as above for the faint magnitude cutoff\footnote{The latter
is adequate due to the much denser spacing of magnitude bins at
fainter magnitudes}; and (2) adopt a slightly smaller bin size of 200
GCs. We then vary the upper cutoff in steps of 0.25 mag between
$M_z=-12.5$ mag an $M_z=-8.5$ mag.  For each restricted sample, we fit
red and blue \kmm peaks. In the bottom left panel, we overplot the
\kmm peak results for all adopted bright GC magnitude limits. For the
{\tt high-mass} sample, the peak positions of the blue GC population
``bend" slightly at higher GC masses.  In the bottom right panel of
Fig.~\ref{slopes}, this behavior is quantified by plotting $\gamma_z$
as a function of the upper cutoff magnitude. For GCs belonging in the
{\tt high-mass} sample, the slope is to within the errors independent
of the upper cutoff for $-11<M_{\rm z,cut,bright}<-9$ mag, but
increases somewhat from $\gamma_z=-0.040$ to $\gamma_z=-0.052$ when
including also the brightest GCs with $-12.5<M_z<-11$ mag. For GCs in
the {\tt low-mass} sample, no steepening for bright GCs is apparent.

In summary, the tilt becomes stronger for the most massive GCs
($\gtrsim 2-3 \times 10^6$M$_{\sun}$) belonging to the {\tt high-mass}
sample, but not for the GCs in the {\tt low-mass} sample.

\subsection{Evidence for a Red Tilt?}

We now examine the evidence for a ``red tilt" in the data: i.e., a
color-magnitude relation for the subpopulation of red
GCs. Table~\ref{Fitresults} shows that there is indeed a 2-3$\sigma$
effect that is detectable in the {\tt high-mass} samples, indicating a
gentle mass metallicity relation of $Z \propto M^{0.1-0.2}$. 
There is, however, no significant red tilt detected in the full sample
of all GCs. Given this and the fact that the colour distribution in
the red peak is generally asymmetric (see Figs. 2 and 5 of Peng
\etal \cite{Peng06}), the evidence for a red tilt should still be
considered preliminary. We note that a weak tilt is broadly
consistent with the GC self-enrichment scenario, where the
pre-enrichment level of red (metal-rich) GCs is significantly higher
than that of the blue (metal-poor) GCs, thereby making the detection
of any self-enrichment more difficult (e.g. ACSVCS XIV, Strader \&
Smith~\cite{Strade08}, Bailin \& Harris ~\cite{Bailin09}).

\vspace{0.4cm}

\noindent We conclude this section with a note of caution regarding
the detection of tilts, red or blue: the colors of individual GCs
typically scatter significantly about the individual peak positions by
0.1-0.2 mag. This scatter -- likely caused by stochastic variation in
the GC pre-enrichment level -- will allow the detection of any
mass-metallicity trend only for systems of at least a few hundred
GCs. The significance of the blue tilt detection in the GC
system of the disk galaxy NGC 5170 (Forbes et al.~\cite{Forbes10}) --
which has about 600 GCs -- is $\sim$3$\sigma$, showing that it would
be difficult to detect subtle tilts in poorer GC systems like that of
the Milky Way (see also Strader \& Smith~\cite{Strade08} and Bailin \&
Harris~\cite{Bailin09}).

\section{Discussion}

In ACSVCS XIV, it was shown that neither stochastic
fluctuations within the stellar CMDs, contamination of the GC samples
by tidally-stripped nuclei, nor the accretion of GCs like those found
in present-day galaxies could fully explain the observed
color-magnitude relations.  Subsequently, in Mieske \&
Baumgardt~\cite{Mieske07}, it was shown through N-body simulations
that field star capture is much too inefficient to account for any
tilt. In this section, we examine the self-enrichment scenario put
forward by Bailin \& Harris~\cite{Bailin09}, including new constraints
imposed on this scenario by the observed trends with environment
(\S4.1).  We then quantify the effects of GC dynamical evolution on
the distribution of clusters within the color-magnitude plane (\S4.2).

\subsection{Constraints on Self-Enrichment Scenarios}

Models for the self-enrichment of star clusters have a long history
(see, e.g., Frank \& Gisler~\cite{Frank76}, Smith~\cite{Smith96},
Gnedin \etal~\cite{Gnedin02}, Parmentier \& Gilmore~\cite{Parmen01},
Dopita \& Smith~\cite{Dopita86}, Morgan \& Lake~\cite{Morgan89}, Thoul
\etal~\cite{Thoul02}, Recchi \& Danziger~\cite{Recchi05},
Parmentier~\cite{Parmen04}, Baumgardt {\it et al.}~\cite{Baumga08})
and they appear to offer a promising explanation for the blue
tilt.  The mass-metallicity relations as reported
in this paper range from $Z \propto M^{0.3}$ to $\propto
M^{0.7}$. This overlaps with the $Z \propto M^{0.6-0.7}$ relations
found for dwarf spheroidal galaxies (e.g., Simon \&
Geha~\cite{Simon07}, Kirby {\it et al.}~\cite{Kirby08}) which have
luminosities, metallicities, and central velocity dispersions that are
similar to those of GCs (although the latter appear to be baryon,
rather than dark matter, dominated).

In Strader \& Smith ~\cite{Strade08}, Bailin \& Harris~\cite{Bailin09}
and Harris {\it et al.}~\cite{Harris09b} the self-enrichment scenario
of star clusters is discussed with a focus on the blue tilt. The
underlying assumption is that self-enrichment by SNII ejecta starts to
become efficient when the energy output by SNII is comparable to the
binding energy of the cluster's primordial gas cloud.  This implicitly
assumes that 100\% of the SN ejecta energy is converted to kinetic
energy of the cloud gas (see Tenorio-Tagle {\it et
al.}~\cite{Tenori07} for an alternative view).  Bailin \&
Harris~\cite{Bailin09} parameterize the ability for self-enrichment of
a proto-cluster cloud based on the metal-retention fraction, $f_Z$
(equation 7 of their paper),
\begin{equation}
\log Z_c/Z_{\odot} = \log (10^{Z_{pre}} + 10^{0.38 + \log f_* f_Z}),
\label{Ztot}
\end{equation}
where $f_*$ denotes the star formation efficiency. The particular
value of 0.38 used in the exponent of the second term is determined by
the SN yield. For the metal-retention fraction, Bailin \& Harris
derive (equation 28 of their paper):

\begin{equation}
f_Z = \exp( -\frac{E_{SN} f_* r_t}{10^2~M_{\odot} G M_c})
\label{fzist}
\end{equation}

This expression holds for the assumption of a 3D mass density profile of the
protocluster cloud that falls off as $\rho \propto r^{-2}$ (i.e., an isothermal
sphere), with $r_t$ being the truncation radius. In this equation,
$E_{SN}$ is the typical energy released by a single SNII ($=10^{51}$
erg) and $M_c$ is the mass of the proto-cluster cloud. A Salpeter-like
stellar mass function at the high-mass end is assumed. [Note that the
present-day cluster mass is linked to the proto-cluster cloud mass via
the relation
$M_{c,today}=M_c f_*$.]
For a more general density profile
$\rho \propto r^{-\beta}$ with $\beta \neq 2$, it holds for $f_Z$ (equation 36 of
their paper):

\begin{equation}
f_Z = [ 1 - \frac{E_{SN} f_* r_t
    (2-\beta)}{10^2~M_{\odot} G M_c }]^{\frac{3-\beta}{2-\beta}}
\label{fzbeta}
\end{equation}

\begin{figure*}
\centerline{\plottwo{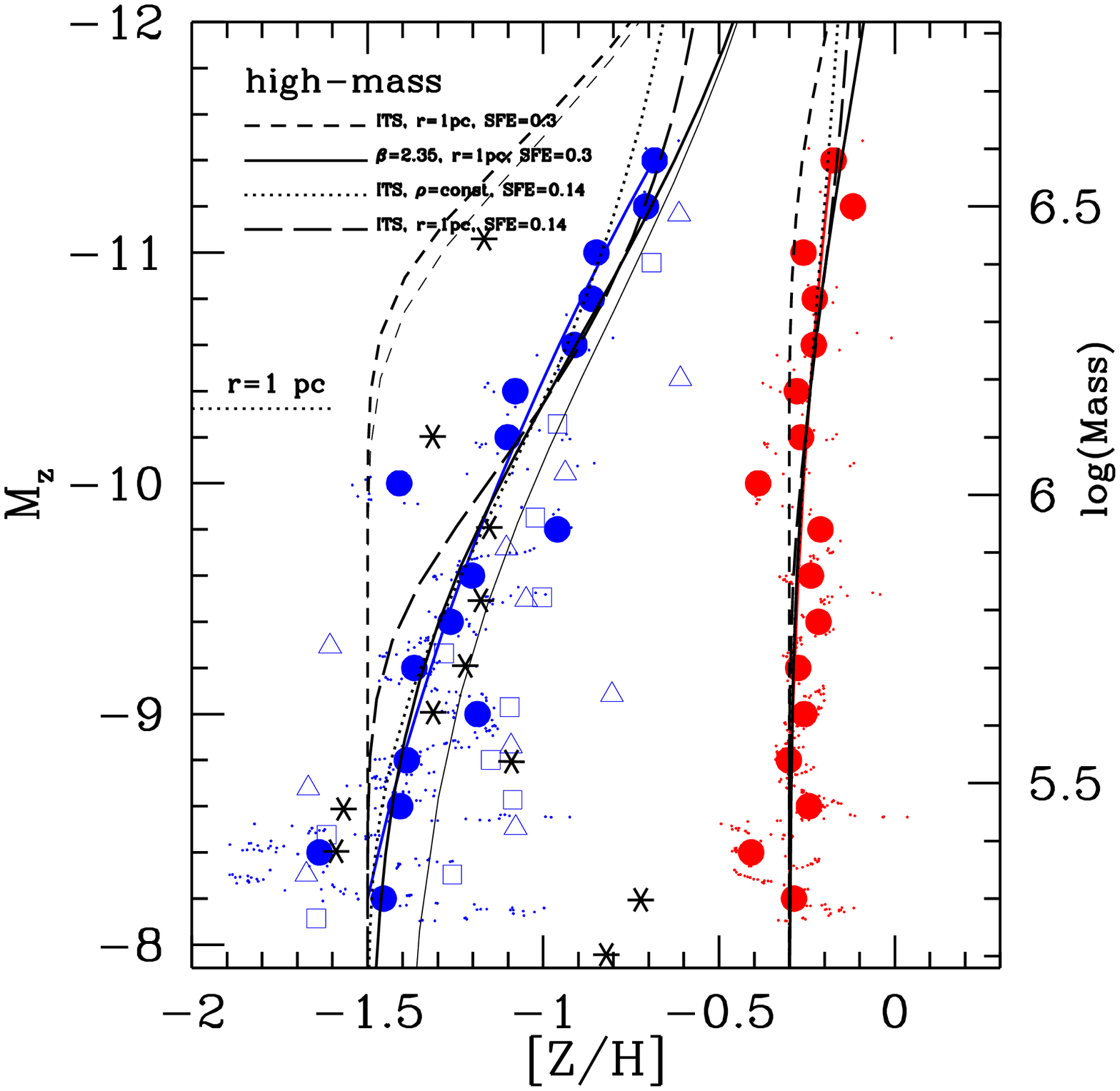}{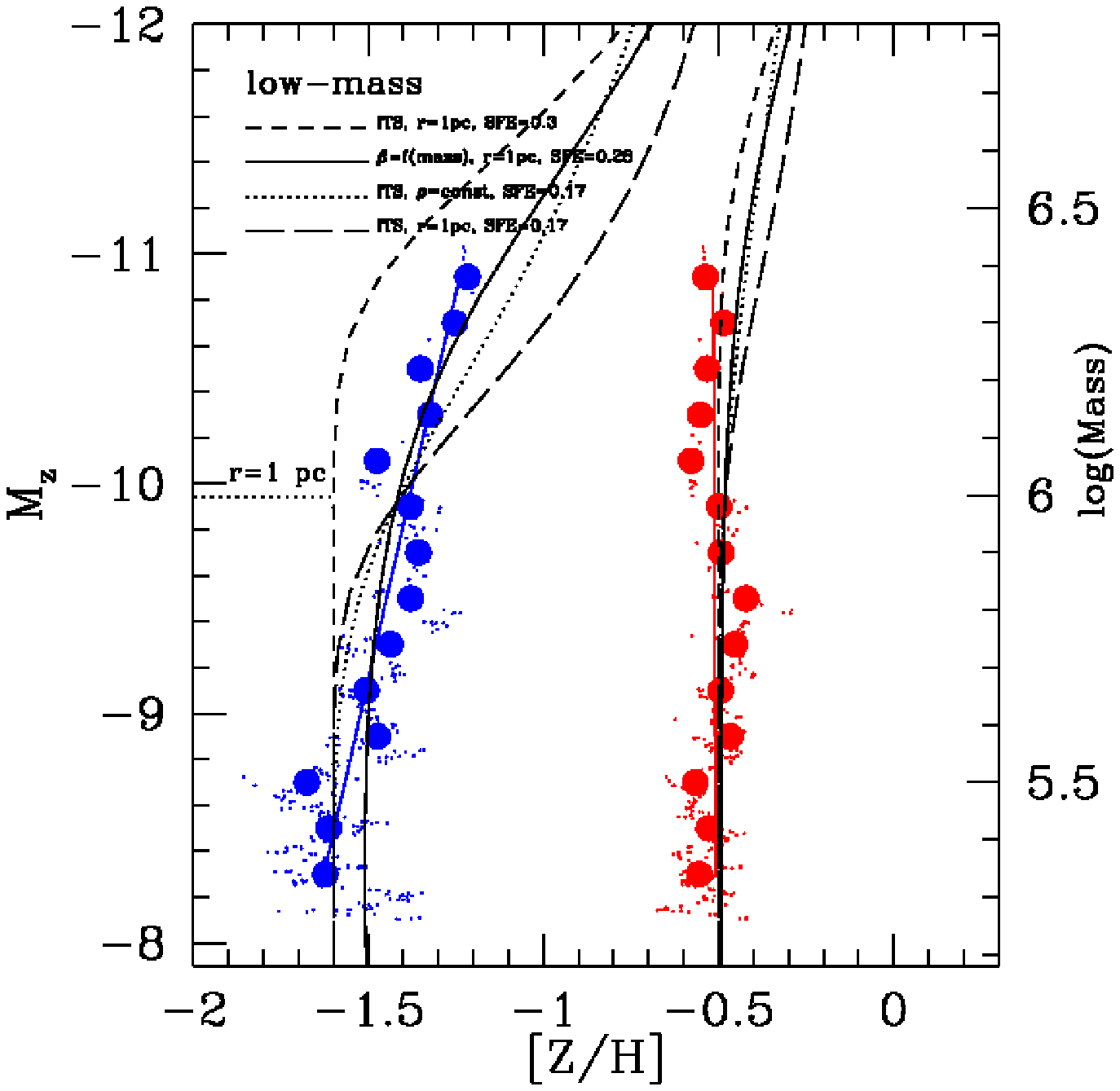}}
\centerline{\plottwo{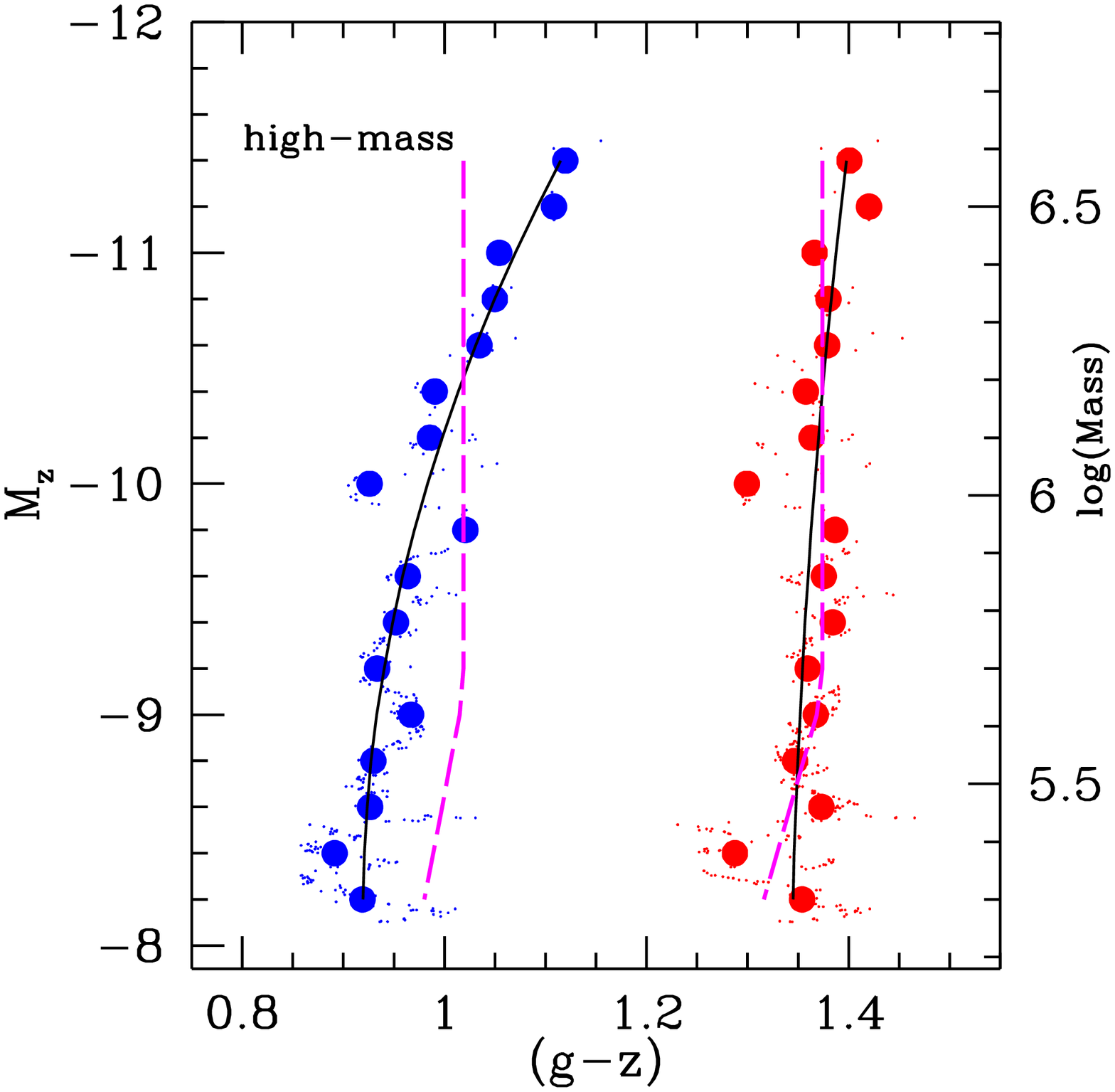}{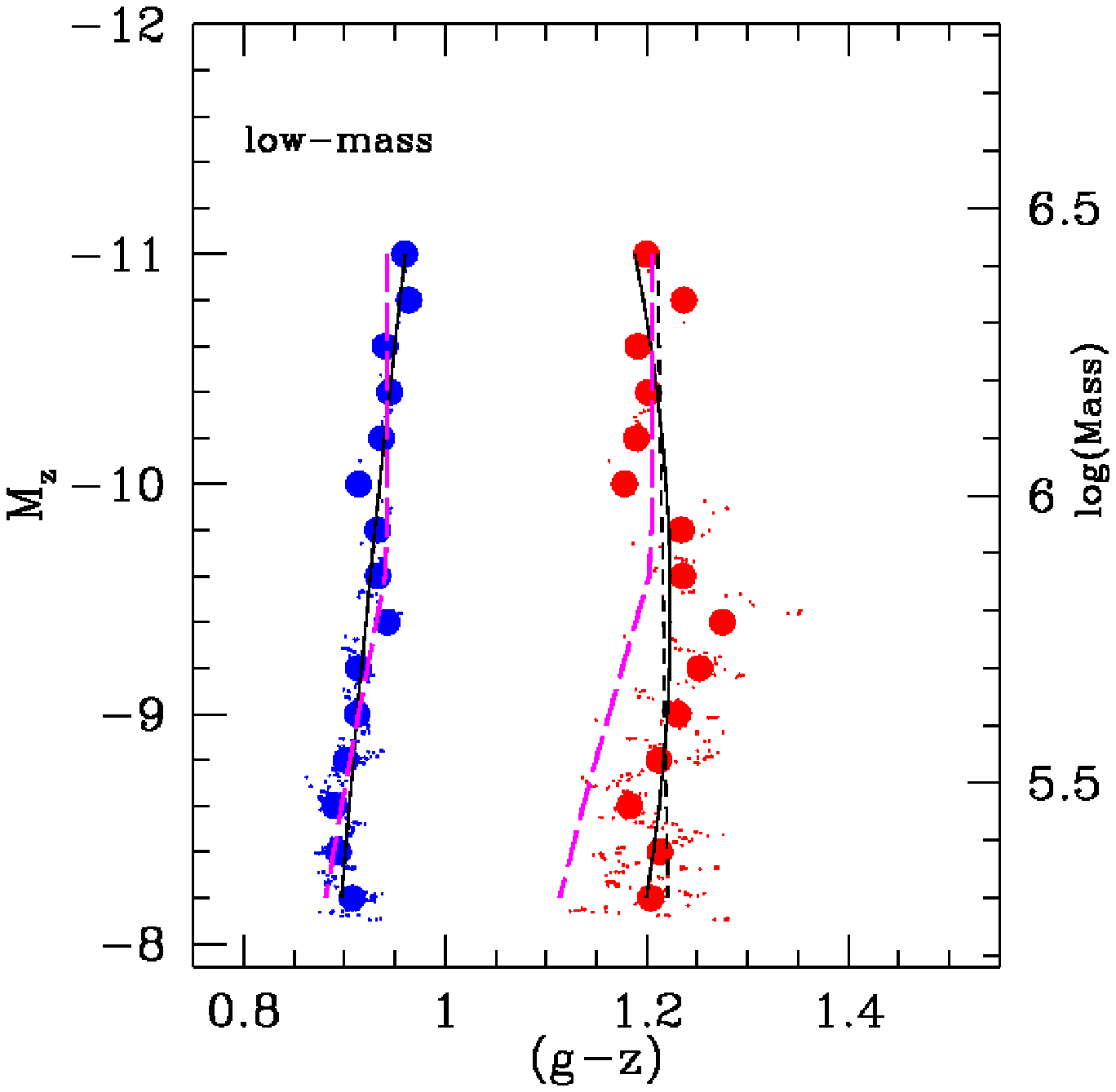}}
\caption{\label{CMDphotev}\label{sem}Comparison of the observed tilt
(blue and red) to predicted changes of metallicity/color caused by
self-enrichment {\it (top panels)} and dynamical evolution {\it
(bottom panels)}. We consider separately the GC subsamples belonging
to {\tt high-mass} and {\tt low-mass} host galaxies. The filled blue
and red solid circles correspond to the mean GC metallicity/color in
magnitude bins of 0.2 mag also shown in the bottom left panel of
Fig.~\ref{slopes}. The colors $(g-z)$ have been converted to
metallicity [Z/H] using the biquadratic transformation of Peng et
al. ~\cite{Peng06}. In the top left panel, the open triangles indicate
the individual peak positions for NGC~1399, the squares indicate M87,
and the asterisks show M49.  {\it (Top Panels: A Test of
Self-Enrichment).} The solid blue and red curves indicate quadratic
fits to the \kmm data points while the black curves are predictions
from the self-enrichment model of Bailin \& Harris~\cite{Bailin09} for
various sets of model parameters, as indicated in the plot legend. The
reference model adopted by Bailin \& Harris~\cite{Bailin09} is
indicated by the short-dashed curve. This particular model predicts
self-enrichment to appear at an order of magnitude larger mass than is
observed. The thin short dashed and thin solid line are a
variation of the respective thick line models, assuming a constant
mass loss of $\Delta=3 \times 10^5M_{\odot}$ to have occured for all
clusters. {\it (Bottom Panels: A Test of Dynamical Evolution).} The
solid curves are quadratic fits to the observed color-magnitude data
points. The long dashed magenta curves indicate the expected color
change due to dynamical evolution according to the predictions from
Kruijssen \& Lamers (2008). For the red peak at {\tt low-mass}
galaxies the short dashed line also shows a linear fit to the plotted
data points.}
\end{figure*}

Equations~\ref{fzist} and~\ref{fzbeta} show that the primordial star
formation efficiency, $f_*$, the proto-cluster mass-radius relation,
and density profile, $\rho$, will determine the self-enrichment
capability of a given cluster.  Our aim is to constrain these
parameters based on the observed GC mass-metallicity relation. To this
end, we plot in the top panels of Fig.~\ref{CMDphotev} GC metallicity
vs. absolute magnitude and mass, separated into {\tt high-mass} and
{\tt low-mass} samples. The [Z/H] values are obtained directly from
converting the $(g-z)$ datapoints to [Fe/H] using the biquadratic
transformation based on Peng \etal~\cite{Peng06}. We also show separately
the data for the three high-luminosity galaxies NGC~1399, M49 and M87
(see \S3.2.1).

Fig.~\ref{CMDphotev} shows the expected behaviour of $\log
Z_c/Z_{\odot}$ as a function of mass for various choices of model
parameters. The short dashed curve corresponds to the reference model
adopted by Bailin \& Harris~\cite{Bailin09} (see also Harris et
al.~\cite{Harris09b}). For this model, the density profile is taken to
be an isothermal sphere with a mass-independent truncation radius of
$r_t=1$~pc, and a star formation efficiency of $f_*=0.3$ (e.g. Boily
\& Kroupa~\cite{Boily03}, Lada \& Lada~\cite{Lada03}, Baumgardt {\it et
al.} 2008).

This particular set of model parameters does not reproduce the data
for either the {\tt high-mass} nor the {\tt low-mass} sample. Most
notably, the observations point to an onset of the blue tilt at
$\approx 2 \times 10^5 M_{\odot}$, while the reference model predicts
self-enrichment to become important only above $\approx 2 \times 10^6
M_{\odot}$. Moreover, the relation between mass and metallicity
predicted by the model in the self-enrichment mass regime is $Z\propto
M^{1.0-1.1}$. This is significantly steeper than the observed relation
of $Z\propto M^{0.3-0.7}$.

Thus, the data require a lower mass for the onset of self-enrichment
than in the reference model, but at the same time, a weaker scaling
between metallicity and mass in the self-enrichment regime. One way
of achieving this is to assume a radial density profile that is
steeper than an isothermal sphere: Fig.~\ref{CMDphotev} shows that the
particular choice of $\beta=2.35$ matches very well the observed
mass-metallicity scaling for the {\tt high-mass} sample. For such a
profile, self-enrichment is able to start at lower cloud masses due to
the deeper potential well in the cloud center, but its efficiency
increases more gradually with increasing cluster mass (given that only
a small fraction of gas is located close to the edge of the
proto-cluster cloud). For the {\tt low-mass} sample, the data require
a variation of $\beta$ with cloud mass in the sense that lower-mass
clouds have a steeper profile than higher-mass clusters; the shallower
tilt for this sample can otherwise not be reproduced. The solid curve
in Fig.~\ref{CMDphotev} corresponds to the particular choice of
$\beta=2.3 + 0.23[6.2-\log(\rm mass)]$.

An alternative way to match the observations for the {\tt high-mass}
sample is to adopt a lower star formation efficiency of $f_* \simeq
0.15$. This implies that the progenitor cloud for a GC of a given mass
was twice as massive as in the reference model, yielding a
correspondingly larger metal retention fraction, and hence, a lower
{\it present-day} onset mass for self-enrichment. At the same time,
the choice of a lower $f_*$ reduces the overall efficiency of self
enrichment, that is, the slope between [Z/H] and mass
(Equ.~\ref{Ztot}). For the {\tt low-mass} sample, however, the curves
with lower star formation efficiency do not reproduce the observed
scaling of metallicity with mass.

It is important to bear in mind that the assumption of a
constant cloud radius is not necessary to reproduce the
observations. Fig.~\ref{CMDphotev} shows that clouds with constant
density (i.e., $r \propto M^{1/3}$) can also have levels of
self-enrichment that are in agreement with the observations. 

Finally, we note that radiative feedback from massive stars also
contributes to regulating star formation, with the total radiative
energy input from OB stars into the ISM being more than an order of
magnitude lower than the energy injected by SNII (Baumgardt et
al.~\cite{Baumga08}). Despite the lower amount of energy involved,
radiative feedback may be an important factor for halting star
formation in massive clusters (e.g. Krumholz \&
Matzner~\cite{Krumho09}, Fall et al.~\cite{Fall10}) in the few million
years before the first SNII explode. This effect is not considered in
the self-enrichment scenario discussed in the present paper, since we
focus on the reprocessing of metals ejected by SNII, which dominate
the budget of metal-enriched material in a star cluster.

\subsubsection{Environmental Dependence}

We have shown that the mass-metallicity scaling relation, and the
difference between GCs in {\tt high-mass} and {\tt low-mass} samples,
may be explained if the density profiles of the proto-cluster clouds
are generally steeper than isothermal spheres, and if the density
profiles for the proto-cluster clouds in the {\tt low-mass} sample
depended weakly on mass. Another effect that may contribute to the
observed trends is that, for larger cluster masses, an increasing
fraction of proto-cluster clouds with low star formation efficiency
survive their formation (Baumgardt {\it et al.}~\cite{Baumga08}). This is
possible since, for higher-mass clusters, gas expulsion will tend to
happen more adiabatically and the tidal radius is larger, so that
clusters will have more room for expansion folllowing gas
expulsion. This possibility is also supported by the finding of
Baumgardt {\it et al.}~\cite{Baumga08} that, in order to turn a power-law
cluster mass function into the commonly observered log-normal one
(e.g., Jord\'an \etal\ \cite{Jordan07b} and references therein), a larger
fraction of high mass clusters has to survive gas expulsion.
Environmental variations in the importance of these effects are
certainly conceivable: e.g., in the sense that metal retainment is
favoured in the central parts of high-mass galaxies with external
pressure confinement, different from regions in lower mass galaxies
with a weaker external pressure (e.g., Parmentier~\cite{Parmen04},
Recchi \& Danziger~\cite{Recchi05}). These variations may well
influence the resulting mass-metallicity relation of GCs.

However, such smooth trends can hardly explain the gross differences
between the GC CMD morphology for galaxies of similar morphological
type and luminosity, as is the case for M49, M87 and NGC~1399
(Fig.~\ref{3giants} and Fig.~\ref{CMDphotev}). Instead, these
differences may indicate that individual merger or accretion events
have altered their GC populations (e.g., C\^{o}t\'{e}
\etal~\cite{Cote98} or Hilker \etal~\cite{Hilker99}), and hence
contributed to modulating the blue tilt. This modulation might take
the form of intermediate-age GCs formed in gas-rich mergers, the
dissipationless accretion of GCs belonging to gas-poor galaxies, or
both (see the reviews of West \etal \cite{West04} and Brodie \&
Strader~\cite{Brodie06}).

An alternative explanation for the variations in the color-magnitude
relations among different subsamples is considered by Blakeslee {\it
et al.}~\cite{Blakes10}. These authors show that the non-linear
relationship between GC color and metallicity may lead to different
tilts in color space between samples of different mean metallicity,
even though the underlying GC mass-metallicity relations may be quite
similar. In the present paper we adopt a color-metallicity tranformation
based on  95 GCs in the Milky Way, M49 and M87 (Peng {\it et
al.}~\cite{Peng06}), the same data set which is also adopted for the
colour-metallicity transformation by Blakeslee {\it et
al.}~\cite{Blakes10}. Since we do see a difference in the
mass-metallicity relation between GCs in {\tt high-mass} and {\tt
low-mass} galaxies, not only in color-magnitude space, it appears
likely that the effect considered by Blakeslee {\it et al.} is in our
case not the only cause for the environmental variation. An important
step forward in this context would be to obtain accurate spectroscopic
metallicity measurements for an expanded sample (e.g., Cohen \etal\
\cite{Cohen98} and \cite{Cohen03}) of GCs in the Virgo and Fornax
galaxies that define the tilt and its variations.

\subsection{Color Effects from Dynamical Evolution}

Having constrained the conditions under which self-enrichment can
reproduce the observed blue tilt, it is worth exploring the extent to
which colour changes due to dynamical evolution of GCs may also contribute to shaping the
observed trends. In the context of the environmental variations of the
tilt, this issue is worth considering in some more detail since the
dynamical evolution of star clusters is known to depend strongly on
the external tidal field and hence varies as a function of
environment.

Two recent studies investigate the effects of tidal dissolution of
star clusters on their integrated colors (Kruijssen \&
Lamers~\cite{Kruijs08}; Anders {\it et al.}~\cite{Anders08}), based on
the N-body star cluster simulations of Baumgardt \& Makino
~\cite{Baumga03}. Of these two studies, Anders {\it et
al.}~\cite{Anders08} perform a more sophisticated treatment of the
evolution of the stellar mass function change with time, resulting in
somewhat smaller predicted color changes.  Both studies predict star
cluster colors at a determined age and metallicity as a function of
the ratio t$_{dyn}$ between current age and dissolution
time. Generally, color changes start to set in after $\sim$0.4
dissolution time scales.

To determine whether these changes can be responsible for the tilt, we
estimate the amount of color change that has occured in the GCs of the
combined VCS and FCS sample. We adopt a $z$-band mass-to-light ratio
of ${M}/{L_z}=1.5$, independent of $(g-z)$ color (Jord\'an et
al.~\cite{Jordan07b}). As the fractional dynamical age, $t_{dyn}$, we
define the ratio of the current age to the dissolution time. In the
context of the GC dissolution scenario discussed in Jord\'an {\it et
al.}~\cite{Jordan07b}, this can also be expressed as the ratio between
present-day and primordial mass. Using the present-day mass,
$M_{present}$, and the accumulated mass loss, $\Delta$, we find:

\begin{equation}
t_{dyn}=1-\frac{M_{present}}{M_{present}+\Delta}
\label{delta}
\end{equation}

We assume $\Delta$ to be equal for all GCs in a given host galaxy,
adopting $\Delta=3 \times 10^5M_{\odot}$ for {\tt high-mass} galaxies
and $\Delta=5 \times 10^5M_{\odot}$ for {\tt low-mass} galaxies (see,
e.g., Fig.~16 of Jord\'an {\it et al.}~\cite{Jordan07b}). In the bottom
panels of Fig.~\ref{CMDphotev} we show the \kmm peak positions fitted
to the CMDs of GCs from the combined FCS and VCS {\tt high-mass} and
{\tt low-mass} samples. We also show quadratic fits to
the \kmm peak positions. In each luminosity bin we calculate the mean
GC mass, the mean host galaxy magnitude and adopt $\Delta$ in
equation~\ref{delta} accordingly, yielding an average $t_{dyn}$ for
the GCs in each luminosity bin. We then calculate an estimate for the
color change at each luminosity bin, adopting the predictions by
Kruijssen \& Lamers ~\cite{Kruijs08} in the $V, I$ bands as a function
of $t_{dyn}$. We use ${\delta (V-I)}/{\delta (g-z)}=0.521$ (Peng et
al. 2006) to convert to $(g-z)$ colors.  The results are indicated by
the long-dashed magenta lines. It is apparent that non-negligible
color changes are expected for only the low-luminosity GCs: for $M_z
\gtrsim -9.0$ mag in the {\tt high-mass} sample, and $M_z \gtrsim
-9.5$ mag for the {\tt low-mass} sample. Moreover, the predicted
strong changes in the red peak for the {\tt low-mass} do not at all
agree with the observational data. This comparison thus supports the
relatively small color changes predicted by Anders et
al.~\cite{Anders08}; we therefore conclude that dynamical evolution
may be responsible for a weak color-magnitude trend among low-mass
GCs, but certainly not for the strong tilt exhibited by the more
massive clusters.

We also note that the mass-loss of star clusters only has a very
small effect on the self-enrichment estimates presented in the
previous Section. This is illustrated by two example self-enrichment
model calculations in the upper left panel of
Fig.~\ref{CMDphotev}. The thin short dashed and thin solid line show
the respective thick line model variations assuming a constant mass loss
of $\Delta=3 \times 10^5M_{\odot}$ to have occured for all
clusters. The effect on the reference model by Bailin \& Harris is
negligible, and the effect for the model variations that fit the data
corresponds to a 10-15\% change in slope. A 0.15 dex lower
pre-enrichment metallicity is required for the predictions that include mass-loss to match the predictions without mass-loss.

\section{Conclusions}
A color-magnitude relation for metal-poor GCs --- the ``blue tilt" ---
is now a well established observational result. From a CMD analysis 
based on a sample of $\approx$ 17000 GCs detected in the ACS Virgo and
Fornax Cluster Surveys, we identify a clear environmental variation in
the slope of the GC color-magnitude relation: i.e., the slope of the
color-magnitude relation is largest for the highest-mass
galaxies. Nevertheless, we confirm that there are real,
galaxy-to-galaxy differences in the amount of tilt at comparable
galaxy luminosity and mass: e.g., M87 and NGC~1399 show significant
blue tilts, while M49 does not (Mieske \etal\ 2006, Strader \etal\
2006).  

A comparison to predictions from models of GC self-enrichment (Bailin
\& Harris~\cite{Bailin09}) suggest that this mechanism offers a
promising explanation for the observed relation and may be able to
constrain the conditions under which star formation took place in
proto-cluster clouds. Within the context of this model, the observed
shape of the color-magnitude relation suggests that protocluster
clouds had density profiles somewhat steeper than isothermal in the
highest-mass galaxies ($\beta \approx 2.3$). For the lower mass
galaxies, a weak mass dependence for the structure of the protocluster
clouds is required to match the observations (with $\beta$ in the
range $2.5 \gtrsim \beta \gtrsim 2.2$ for masses $10^{5.5} \lesssim
M/M_{\odot}\lesssim 10^{6.5}$). This interpretation, however, is not
unique. Other variants on this same basic picture also appear
consistent with the observations, including models in which the clouds
have constant density ($r \sim M^{1/3}$) and/or relatively low star
formation efficiencies 0.15-0.20.

However, we point out that the significant differences in the detailed
CMDs for some galaxies of similar mass and morphological type pose a
challenge to single generic mechanisms, such as self-enrichment, that
attempt to explain GC color-magnitude relations. These differences
suggest that the detailed (and stochastic) merger/accretion histories
of individual galaxies will play a role shaping the appearance overall
GC CMDs.

We conclude that further insights into the blue tilt phenomenon now
await optical spectroscopy for large, representative GC samples
covering the upper 3-4 magnitudes of the GC luminosity function, where
the color-magnitude relation is most apparent.  At the distances of
the Fornax and Virgo clusters, this corresponds to apparent magnitudes
of $21 \lesssim V \lesssim 24.5$~mag, just within the reach of
existing 8-10m-class telescopes.

\acknowledgements 
Support for programs GO-9401 and GO-10217 was provided through grants from
STScI, which is operated by AURA, Inc., under NASA contract NAS5-26555. AJ. and LI 
acknowledge support from the Chilean Center of Excellence in Astrophysics and Associated 
Technologies and from the Chilean Center for Astrophysics FONDAP 15010003. 
MJW. acknowledges support through NSF grant AST 02-05960.
This research has made use of the NASA/IPAC Extragalactic Database (NED) which is 
operated by the Jet Propulsion Laboratory, California Institute of Technology, under contract 
with the National Aeronautics and Space Administration.

\medskip

Facilities: HST(ACS/WFC)

{}

\begin{table}[hb!]
\caption{Color-magnitude trends for red and blue globular clusters in 
ACS Fornax and Virgo Cluster Survey galaxies, determined with \kmm fits\label{Fitresults}}
\tiny
\begin{center}
\begin{tabular}{|l||l|r||l|r||}
\hline\hline Sample &  $\gamma_{z,\rm blue}$ &  n ($Z \propto M^{\rm n}$)& $\gamma_{z,\rm red}$ & n ($Z \propto M^{\rm n}$)\\\hline
NGC 1399 ($M_B=-20.9$)&              $-$0.0878 $\pm$ 0.0250 [$-$0.0863]$^*$& 0.82 $\pm$ 0.23 &$-$0.0256 $\pm$ 0.0050 [$-$0.0215]$^*$& 0.31 $\pm$ 0.12\\
FCS {\tt high-mass} ($-22.1<M_B<-19.75$)&          $-$0.0697 $\pm$ 0.0116 [$-$0.0737]$^{**}$ & 0.70 $\pm$ 0.12 &$-$0.0363 $\pm$ 0.0091 [$-$0.0309]$^{**}$ & 0.25 $\pm$ 0.07 \\
FCS {\tt low-mass} ($-19.75<M_B<-16$)&        $-$0.0186 $\pm$ 0.0066 [$-$0.0177]& 0.28 $\pm$ 0.10 &$+$0.0101 $\pm$ 0.0125 [$+$0.0154]& $-$0.04 $\pm$ 0.05\\
FCS                  &             $-$0.0257 $\pm$ 0.0050 [$-$0.0215]& 0.38 $\pm$ 0.08 &$-$0.0078 $\pm$ 0.0117 [$-$0.0033]& 0.03 $\pm$ 0.05\\\hline 
VCS   &                            $-$0.0459 $\pm$ 0.0048 [$-$0.0352]& 0.62 $\pm$ 0.08 &$-$0.0223 $\pm$ 0.0095 [$-$0.0026]& 0.12 $\pm$ 0.05 \\
FCS+VCS {\tt high-mass} ($-22.1<M_B<-19.75$)&           $-$0.0566 $\pm$ 0.0076 [$-$0.0514]& 0.66 $\pm$ 0.09 &$-$0.0198 $\pm$ 0.0099 [$-$0.0123]& 0.10 $\pm$ 0.05 \\
FCS+VCS {\tt low-mass} ($-19.75<M_B<-15.4$)&          $-$0.0210 $\pm$ 0.0034 [$-$0.0231]& 0.33 $\pm$ 0.06 &$-$0.0076 $\pm$ 0.0087 [$-$0.0072]& 0.03 $\pm$ 0.03\\
FCS+VCS {\tt high-mass}, d$<5.5$ kpc &          $-$0.0750 $\pm$ 0.0120 [$-$0.0700]& 0.76 $\pm$ 0.12 &$-$0.0301 $\pm$ 0.0094 [$-$0.0225] & 0.20 $\pm$ 0.07\\
FCS+VCS {\tt high-mass}, d$>5.5$ kpc &          $-$0.0428 $\pm$ 0.0071 [$-$0.0267]& 0.58 $\pm$ 0.10 &$-$0.0094 $\pm$ 0.0120 [$-$0.0041]& 0.04 $\pm$ 0.06\\
FCS+VCS&                          $-$0.0293 $\pm$ 0.0085 [$-$0.0312]& 0.43 $\pm$ 0.12 &$-$0.0082 $\pm$ 0.0190 [$-$0.0034]& 0.03 $\pm$ 0.08\\\hline\hline
\end{tabular}
\end{center}
\small
\vspace{0.2cm} Notes: For the samples in Column 1 (see
Figs.~\ref{CMDfornax} and~\ref{CMDfv} for the respective CMDs),
Columns 2 and 4 give the slopes $\gamma$ between $(g-z)$ and the $z$-band magnitudes
of the blue and red GC subpopulations, determined from linear fits to the
\kmm peak positions.  Errors are based on random resampling of the data
points using their measured dispersion around the fit. Values in
square brackets are the slopes determined from median colors blueward
and redward of the dividing line between both populations. Columns 3 and 5
give the corresponding scaling relation in mass-metallicity space. The lower
limit used in the  fitting was $M_z=-8.1$ mag for all samples.\\
$^*$ Slopes from median colors at magnitude independent color
division: $\gamma_{z,\rm blue}$=$-$0.0551, $\gamma_{z,\rm
red}$=$-$0.0009 (see Fig.~\ref{CMDfornax}).\\ $^{**}$ Slopes from median
colors at magnitude independent color division: $\gamma_{z,\rm
blue}$=$-$0.0331, $\gamma_{z,\rm red}$=$-$0.0003 (see
Fig.~\ref{CMDfornax}).\\ \normalsize
\end{table}
\end{document}